\documentclass[11pt]{article}
\usepackage{amsmath,amsfonts}
\usepackage{graphicx,color}
\def\sloppy{\tolerance=100000\hfuzz=\maxdimen\vfuzz=\maxdimen}
\def \beq  {\begin{equation}}
\def \eeq  {\end{equation}}
\def \beqar {\begin{eqnarray}}
\def \eeqar {\end{eqnarray}}
\def\sqr#1#2{{\vcenter{\vbox{\hrule height.#2pt
\hbox{\vrule width.#2pt height#1pt \kern#1pt
\vrule width.#2pt}\hrule height.#2pt}}}}

\def\la {{\langle}}
\def\ra {{\rangle}}
\def\vx {{\vec x}}
\def\vy {{\vec y}}

\def\vf {{\varphi}}

\def\bvf {{\bar \varphi}}

\def\Tr {{\rm Tr}}

\def\bp {\bar p}

\def\bA {\bar{A}}

\def\bx {\bar{x}}
\def\by {\bar{y}}

\def\vx {{\vec x}}
\def\vz {\vec{z}}
\def\vy{\vec{y}}
\def\vv {\vec{v}}
\def\vu {\vec{u}}
\def\vw {\vec{w}}

\def\del {\partial}
\def\bdel{\bar{\partial}}

\def\e {\epsilon}
\def\d {\delta}
\def\s {\sigma}

\def\bz {{\bar{z}}}

\def\G {{\cal G}}

\def\bV{{\bar V}}

\def\vf {{\varphi}}
\def \bvf {{\bar \varphi}}

\def\half{\textstyle{1\over 2}}

\begin{document}
\fontfamily{cmr}\fontsize{11pt}{17.2pt}\selectfont
\def \CMP {{Commun. Math. Phys.}}
\def \PRL {{Phys. Rev. Lett.}}
\def \PL {{Phys. Lett.}}
\def \NPBProc {{Nucl. Phys. B (Proc. Suppl.)}}
\def \NP {{Nucl. Phys.}}
\def \RMP {{Rev. Mod. Phys.}}
\def \JGP {{J. Geom. Phys.}}
\def \CQG {{Class. Quant. Grav.}}
\def \MPL {{Mod. Phys. Lett.}}
\def \IJMP {{ Int. J. Mod. Phys.}}
\def \JHEP {{JHEP}}
\def \PR {{Phys. Rev.}}
\def \JMP {{J. Math. Phys.}}
\def \GRG{{Gen. Rel. Grav.}}
\begin{titlepage}
\null\vspace{-62pt} \pagestyle{empty}
\begin{center}
\rightline{CCNY-HEP-09/3}
\rightline{June 2009}
\vspace{1truein} {\Large\bfseries
The Hamiltonian Approach to Yang-Mills (2+1):}\\
\vskip .1in
{\Large \bfseries An Expansion Scheme and Corrections to String Tension}\\
\vskip .1in
{\Large\bfseries ~}\\
{\large DIMITRA KARABALI$^a$},  {\large V.P. NAIR$^b$ and ALEXANDR YELNIKOV$^b$}\\
\vskip .2in
{\itshape $^a$Department of Physics and Astronomy\\
Lehman College of the CUNY\\
Bronx, NY 10468}\\
\vskip .1in
{\itshape $^b$Physics Department\\
City College of the CUNY\\
New York, NY 10031}\\
\vskip .1in
\begin{tabular}{r l}
E-mail:&{\fontfamily{cmtt}\fontsize{11pt}{15pt}\selectfont dimitra.karabali@lehman.cuny.edu}\\
&{\fontfamily{cmtt}\fontsize{11pt}{15pt}\selectfont vpn@sci.ccny.cuny.edu}\\
&{\fontfamily{cmtt}\fontsize{11pt}{15pt}\selectfont yelnikov@yahoo.com}
\end{tabular}

\fontfamily{cmr}\fontsize{11pt}{15pt}\selectfont
\vspace{.8in}
\centerline{\large\bf Abstract}
\end{center}
We carry out further analysis of
the Hamiltonian approach to Yang-Mills theory in 2+1 dimensions
which helps to place the calculation of the vacuum wave function and the string tension
in the context of a systematic expansion scheme. The solution of the Schr\"odinger equation is carried out recursively. The computation of correlators is re-expressed in terms of a
two-dimensional  chiral
boson theory. The effective action for this theory is calculated to first order in our expansion scheme and to the fourth order in a kinematic expansion parameter. The resulting corrections to the string tension are shown to be very small,
in the range $-0.3\%$ to $-2.8\%$, moving our prediction closer to the recent lattice estimates.

\end{titlepage}

\pagestyle{plain} \setcounter{page}{2}
\section{Introduction}
A few years ago, we
initiated a Hamiltonian approach to gauge theories in
$(2+1)$ dimensions \cite{KKN1}. In the
$A_0 =0$ gauge, which is appropriate for a Hamiltonian analysis,
the complex
components of the spatial gauge field, viz.,
$A_z,~A_{\bar z}$, were parametrized
as $A_z = -\partial_z M~M^{-1}$, $A_{\bar z} =M^{\dagger -1}\partial_{\bar
z} M^\dagger$, where
$M,~M^\dagger$ are $SL(N,{\mathbb C})$-matrices for an $SU(N)$-gauge theory.
The hermitian matrix $H=M^\dagger M$ then gives the gauge-invariant degrees of freedom.
The Jacobian for the change of variables from $(A_z , A_{\bar z})$ to
$H$ was explicitly calculated. This also led to
the computation of the
volume element on the physical configuration space, and hence
the inner product of wavefunctions, in terms of the
WZW action for the field $H$. The Hamiltonian was then
obtained in terms of the current $J = (N/\pi ) \partial_zH~H^{-1}$
of the WZW action, and the vacuum wave function $\Psi_0$ was calculated
from the Schr\"odinger equation
up to terms which are quadratic in the current $J$ in $\log \Psi_0$.
The vacuum expectation value of the Wilson loop
operator could then be evaluated using this wave functional.
For a Wilson loop in the representation $R$, the result was
 \beq \la W_R(C) \ra  = \exp [- \sigma_{R} {\cal A}_C ]
\label{intro1} 
\eeq 
where ${\cal A}_C$ is the area of the loop $C$. The
string tension $\sigma_R$ was obtained as \cite{KKN2}
\beq
\sqrt{ \sigma_R}  = e^2 \sqrt{ c_A c_R\over
4\pi } \label{intro2} 
\eeq 
where $e$ is the coupling constant, and
$c_R, ~c_A$ denote the quadratic Casimir values for the
representation $R$ and for the adjoint representation, respectively.
This value of the string tension
is in very good agreement with lattice estimates \cite{teper,
bringoltz, kiskis}. 
Our Hamiltonian analysis has also been
extended to the Yang-Mills-Chern-Simons
theory; the wave functions helped to clarify some
issues regarding the dynamical contribution to the gauge boson mass and
screening of fields \cite{nairYMCS}.

Some of the more recent results in this approach include:
\begin{enumerate}
\item A proposal for the wave function, which is very close to ours, but somewhat different, was made by Leigh, Minic and Yelnikov \cite{LMY}.
Using this wave function, an estimate of glueball masses was made.
The values come out to be close to the lattice results with differences
of a few percent. (It should be noted that the lattice values also have significant errors,
especially for the higher glueballs.)

\item The Hamiltonian formalism was extended to include scalar fields \cite{AKN}.
The screening
of Wilson loops in the adjoint and other screenable representations can be related to the formation of a bound state between a heavy scalar field and a light degree of freedom
(the glue part). The energy of this glue-lump bound state (related to the string-breaking point)
was calculated and shown to agree with
lattice estimates to within $\sim 9\%$.
\item The formalism was developed for Yang-Mills theory on ${\mathbb R} \times S^2$
\cite{AN},
motivated by the possibility of connection to gravity-gauge duality \cite{mald}, and also as a first step in developing the method for the torus, i.e., for ${\mathbb R}\times S^1\times S^1$.
The results for the torus can be useful for understanding finite temperature effects and 
deconfinement.

\end{enumerate}

In the light of these results, it is important to formulate a systematic expansion for the wave function and develop a calculational scheme for corrections to
 the string tension. This is the subject of the present paper. The recent lattice calculation of Bringoltz and Teper show that the string tension differs from the prediction
 of (\ref{intro2}) at large $N$ only by about $0.98\% - 1.2\%$ \cite{bringoltz}.
 Nevertheless, the deviation is statistically significant. A different lattice
 method gives a value of string tension which differs from (\ref{intro2}) by
 about $1.55\%$ \cite{kiskis}; again the deviation is considered statistically significant.
 These results provide another motivation for this paper.
 
In terms of an analytical computation, there are two types of corrections to the string tension which may be exemplified by the two graphs shown below, where the
rectangle of solid lines denotes the Wilson loop. The wavy lines represent the propagator
 or two-point 
 \begin{figure}[!b]
\begin{center}
\includegraphics[height = .3\textwidth, width=.65\textwidth]{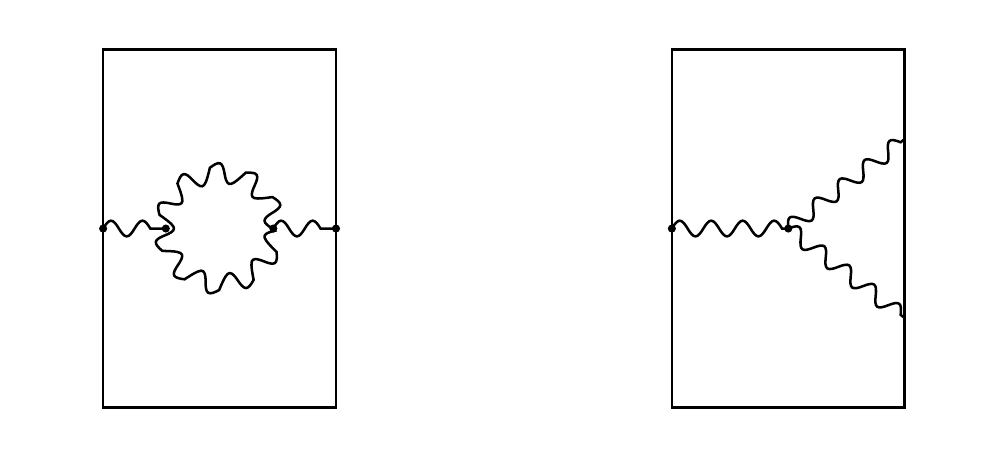}\\
\vskip .1in
\caption{Examples of representation-independent (left) and representation-dependent
(right)
corrections to string tension}
\end{center}
\end{figure}
function for the current $J$.
The first diagram may be viewed as a correction to the propagator.
This correction may be evaluated by computing the corrected propagator in the
sense of an effective field theory. 
(The corrected propagator can then be used for the evaluation of the expectation value of Wilson line.) As a result, it is independent of the representation of the Wilson line.
The second diagram is a correction which cannot be evaluated independently of the Wilson line. It will therefore depend on the representation of the Wilson line.
Both types of corrections are important. 
The second type of corrections can be important in understanding 
screening of Wilson lines and string breaking. 
Some aspects of this problem have been treated elsewhere \cite{AKN}, although a full analysis remains difficult.

In this paper, we will focus on corrections of the first type, namely, those which can be understood as corrections to the propagator. Since the representation-dependent corrections are not considered, the results should be interpreted as applying in the absence of string-breaking, i.e., when all representations have confining area law.
This means that any comparison with lattice value is best done at large $N$.

We start with a short review of the Hamiltonian approach in section 2.
Section 3 gives the basic expansion scheme and
the procedure for calculating higher order corrections. In section 4, we discuss
the calculation of the corrections to the string tension. A conceptual subtle point
is discussed in appendix A and the details of the computation of the various corrections are given in appendix B.
The paper concludes with a short summary and discussion.

\section{A short review}

As is standard in a Hamiltonian analysis, we use the $A_0 =0$ gauge. The spatial components of the gauge potential can be combined as
$A=A_z = \half ( A_1 +i A_2)$, $\bA = A_\bz = \half (A_1 -i A_2)$. These can be parametrized as
\beq
A= - \del M ~M^{-1}, \hskip .2in  \bA = M^{\dagger -1} \bdel M^\dagger .
\label{phi4}
\eeq
$M$ is a complex matrix which is an element of $G^{\mathbb C}$, the complexification of $G$
which is the Lie group in which the gauge transformations take values.
We will be considering $G= SU(N)$, so that $G^{\mathbb C}= SL(N, {\mathbb C})$.

Time-independent  gauge
transformations act on 
${M}$ via ,
\beq {M}(\vec{x}) \rightarrow
g(\vec{x}) {M}(\vec{x})~,  ~~~~         g(\vec{x}) \in SU(N)
\eeq
so that the hermitian matrix $H = M^\dagger M$ is gauge-invariant.

The Jacobian of the tgransformation $A,  \bA \rightarrow H$ can be explicitly evaluated; the volume element for the gauge-invariant configuration space is givenby
\beq
d\mu ({\cal C})
= \int d\mu (H) ~\exp \left( {2c_A S_{wzw}(H)} \right)
\label{sc1} 
\eeq
where $S_{wzw}(H)$ is the Wess-Zumino-Witten action for
the hermitian field $H$ and $d\mu (H)$ is the Haar measure for $H$
viewed as an element of $SL(N, \mathbb{C}) /SU(N)$.
The WZW action is given by
 \beq
S_{wzw} (H) = {1 \over {2 \pi}} \int \Tr (\partial H \bar{\partial}
H^{-1}) +{i \over {12 \pi}} \int \epsilon ^{\mu \nu \alpha} \Tr (
H^{-1}
\partial _{\mu} H H^{-1}
\partial _{\nu}H H^{-1} \partial _{\alpha}H)
\label{sc2}
\eeq

Wave functions are gauge-invariant and have an inner product defined by square-integrability
using (\ref{sc1}).

The kinetic energy operator can be written in terms of its action on functionals of  $H$ as
\beqar
T &=&
-\frac{e^2}{2}\int \frac{\delta ^2}{\delta {A}^a\delta \bar{{A}}^a}  ~
= {e^2 \over 2} \int_{u,v} \Pi_{rs} (\vu,\vv)
\bp_r (\vu) p_s (\vv)\label{phi6} \\
\Pi_{rs} (\vu,\vv) &=& \int_x \bar{\G} _{ar} (\vx,\vu) K_{ab}(\vx) \G _{bs} (\vx,\vv) \nonumber
\eeqar
where $K_{ab} = 2 \Tr (t_a H t_b H^{-1})$ is the adjoint representative of
$H$, and
$p_a$ and 
$\bp_a$ are left and right translation operators for $M^\dagger$ defined by
\beqar
[p_a (\vec{x}), M(\vec{y}) ] &=& M(\vec{y}) (-it_a ) \delta^{(2)}(\vec{x}-\vec{y})\nonumber\\
{}[ \bp_a (\vec{x}) , M^\dagger (\vec{y}) ]  &=& (-it_a) M^\dagger (\vec{y}) \delta^{(2)}(\vec{x}-\vec{y})
\label{phi5}
\eeqar
These may be considered as functional differential operators.
The Green's functions  in (\ref{phi6}) are given by
\beqar
\bar{\G} _{ma} (\vx,\vy)  &=& {1\over \pi (x-y)}   \Bigl[ \d _{ma} - e^{-|\vx-\vy|^2/\e} \bigl(
K(x,\by) K^{-1} (y, \by) \bigr) _{ma}\Bigr] \nonumber\\
\G _{ma} (\vx,\vy)  &=&  {1\over \pi (\bx - \by )} \Bigl[ \d _{ma} - e^{-|\vx-\vy|^2/\e} \bigl(
K^{-1}(y,\bx) K (y, \by) \bigr) _{ma}\Bigr]
\label{phi7}
\eeqar
These are the regularized versions of the corresponding Green's functions
\beq
\bar{G} (\vec{x},\vec{y}) = {1 \over {\pi (x-y)}} ~, ~~~~      G (\vec{x},\vec{y}) = {1 \over {\pi (\bar{x}-\bar{y})}}
\eeq
The parameter controlling the regularization, $\e$, acts as a short-distance cut-off.
Expression (\ref{phi6}) is to be used on functionals where the point-separation
of various factors is much larger than $\sqrt {\e }$.

As discussed in \cite{KKN1}, $(M ,~M^{\dagger})$ and $(M \bV (\bz),~V(z) M^{\dagger})$  give the same gauge potentials $A, ~\bA$, where $\bV,~V$ are, respectively, antiholomorphic and holomorphic in the complex coordinates $\bz = x_1 + ix_2$ and $z= x_1 -i x_2$. To avoid this ambiguity of parametrization, physical observables in the theory should therefore satisfy the holomorphic invariance
\beq
H \rightarrow V(z) H \bV (\bz)
\eeq
The regularization used in (\ref{phi7}) respects this invariance.

Wave functions can be taken to be
functionals of the current $J = (c_A /\pi ) \del H H^{-1}$, where
$c_A$ is the quadratic Casimir invariant defined by
$c_A \delta^{ab} = f^{amn} f^{bmn}$; $c_A =N$ for $SU(N)$.
The action of $T$ on wave functions of the form $\Psi(J)$ can be expressed as
\beq
 T_{YM}~\Psi (J)= {{e^2 c_A} \over {2 \pi}} \left[ \int_z \omega^a(\vz){\delta \over \delta J^a(\vz)}~+\int_{z,w}
\Omega^{ab}(\vz,\vw) {\delta \over \delta J^a(\vz) }{\delta \over
\delta J^b(\vw)}\right] \Psi(J) \label{9} \eeq where \beqar
{}\Lambda _{ra} (\vw,\vz) & = & - \left[ \del _z \Pi_{rs} (\vw,\vz)
\right]
~K^{-1}_{sa} (\vz) \nonumber \\
\omega_a(\vz)& = & if_{arm} \Lambda _{rm} (\vu,\vz)  {\big |}_{\vu
\rightarrow \vz}  \nonumber \\
\Omega_{ab}(\vz,\vw)& = & {\cal{D}} _{w~br} \Lambda_{ra} (\vw,\vz)
\label{10}
\eeqar
with ${\cal{D}}_{w~ab}  = {c_A\over \pi}\partial_w \delta_{ab} +if_{abc}J_c (\vw)$.
For small $\epsilon$, $T$ can be further simplified as
\beq
T_{ YM} \Psi (J) = m \left[ \int J_a (\vz) {\d \over {\d J_a (\vz)}} + \int \bigl(
{\cal{D}} _w \bar{G}(\vz,\vw) \bigr) _{ab} {\d \over {\d J_a (\vw)}} {\d \over {\d
J_b (\vz)}} \right]  \Psi (J) + {\cal O} (\e)
\label{phi9}
\eeq
where $m = e^2 c_A /2\pi$.

The potential energy can be written in terms of the current as
\beq
V_{YM} = {\pi \over {m c_A}} \int d^2x~ :\bdel J^a(x) \bdel J^a(x):
\label{phi10}
\eeq
In terms of  a regularized form with normal ordering, this expression is to be interpreted as
\beqar
V_{YM} &=& {\pi \over {m c_A}} \biggl[ \int_{x,y} \s (\vx,\vy;\lambda ) \bdel J_a (\vx) (K(x,\by)
K^{-1} (y,\by))_{ab} \bdel J_b (\vy) - {{c_A {\rm dim} G} \over {\pi^2
\lambda^{2}}} \biggr] \nonumber\\
\sigma (\vec{x}, \vec{y} ; \lambda) & = & {1 \over {\pi \lambda}} ~\exp\left( -{|\vec{x}-\vec{y}|^2 \over \lambda}
\right)
\label{phi11}
\eeqar
where $\sigma(\vec{x}, \vec{y} ; \lambda)$ is a regularized $\delta$-function, $\lambda$ is the parameter of regularization, and we should take the limit
where $\sqrt{\lambda} \ll 1/e^2$. The operator $U^{ab}(\vx, \vy) = [K(x,\by)
K^{-1} (y,\by)]^{ab}$  is such that the regularized expression for $V_{YM}$ satisfies holomorphic invariance.

The total Hamiltonian, which we shall use in what follows, is thus given by
\beqar
{\cal H} &=& m  \int J_a (\vz) {\d \over {\d J_a (\vz)}} + {m c_A\over \pi^2} \int _{z,w} 
 {1\over (z-w)^2} {\d \over {\d J_a (\vw)}} {\d \over {\d
J_a (\vz)}} \nonumber\\
&&+{\pi \over {m c_A}} \int d^2x~ :\bdel J^a(x) \bdel J^a(x): + i m \int_{z,w} f_{abc} {J^c(w) \over \pi (z-w)} {\d \over {\d J_a (\vw)}} {\d \over {\d
J_b (\vz)}} \label{phi12}
\eeqar

The Schr\"odinger equation for the vacuum wave function was analyzed in 
\cite{KKN1, KKN2} and the leading term in a strong-coupling expansion was obtained as
$\Psi_0 = e^{-{\half} S}$, where
\beqar
S(H) &=&  {{4 \pi ^2} \over { e^2 {c_A}^2}} \int  \bdel J_a \left[ { 1 \over {\bigl( m
+ \sqrt{m^2
-\nabla^2 } \bigr)}} \right] \bdel J_a \nonumber\\
&&\hskip .4in  -2 f_{abc} \int f^{(3)}(\vx,\vy,\vec{z})
J_a(\vx) J_b(\vy) J_c(\vec{z})
+{\cal O}(J^4)
\label{sc2a}
\eeqar
The function $f^{(3)}(\vx,\vy,\vec{z})$ is given in \cite{KKN2}.

Notice that if one restricts to modes of $J$ with momentum $\ll e^2$,
\beqar
S(H) &\approx& {2\pi^2 \over m e^2 c_A^2} \int \bdel J^a \bdel J^a\nonumber\\
&=&{1\over 4 g^2} \int d^2x~ F^a_{ij} F^a_{ij}\label{sc2b}
\eeqar
Computation of expectation values reduces in this limit to a calculation in a Euclidean two-dimensional
Yang-Mills theory with a coupling $g^2 = m e^2$. This was the approximation used in arriving at the formula (\ref{intro2}) for the string tension.

\section{A systematic expansion scheme}

We will start with an outline of our method for calculating corrections to the formula
for the string tensions given in  (\ref{intro2}). 
We shall first rewrite the derivation of the vacuum wave function as a recursive procedure for the solution of the Schr\"odinger equation which will make it clear that (\ref{sc2a}) is the lowest order result
in a systematic expansion.
For this purpose, after a rescaling of the current $J$, we will treat $m$ and $e$ as independent
parameters, setting $m = e^2 c_A/2\pi$ only at the end of all calculations.
In terms of these parameters, $\Psi_0 = e^P$, where $P$ is a power series in $e$.
(However, this is still quite different from perturbation theory since $m$ is included exactly in the lowest order result for $P$. This recursive procedure is something like a resummed perturbation theory. The resummation involves collecting $A, \bA$ in an appropriate series to define $J$ and then including $m$ at the lowest order which is another series.)

The calculation of averages will involve integration of $\Psi_0^* \Psi_0$ over all field space.
Since $J$ is not an independent variable, we will express the integration in terms of a chiral boson field; this transformation is analogous to the fermionization of the WZW model.
We can calculate the corrections to the $J^2$-term (and eventually the string tension) by viewing this version in terms of the chiral boson field as a two-dimensional field theory.
There will be two sets of contributions which are corrections to the $J^2$-term to any given order in $e$.
One term will be a direct contribution from the recursive procedure.
The other set of terms will be loop corrections of the two-dimensional field theory.
These latter terms can be considered as Feynman diagrams. Vertices corresponding to currents in these loops will carry powers of $m/E_k = m/\sqrt{k^2 +m^2}$, where $k$ is some typical momentum.
These factors suppress the contribution of the loop integral, and so, it is advantageous 
to group loop corrections by the number of powers of $m/E_k$.
Our procedure thus involves three steps:
\begin{enumerate}
\item Solving the Schr\"odinger equation as a power series in $e$, after rescaling $J$, and treating $m$ and $e$ as independent parameters;
\item Evaluating the loop corrections in the two-dimensional field theory used for computing expectation values;
\item Grouping loop corrections by powers of $m/E_k$ and calculating all the contributions for each power of $m/E_k$.
\end{enumerate}

\subsection{The recursive solution}

We now turn to the first step in the expansion scheme outlined above, namely, 
the recursive solution of the 
Schr\"odinger equation. For this, we shall rescale the current as
 $J \rightarrow {ec_A/2\pi} J$. The Hamiltonian (\ref{phi12}) now takes the form
 \beqar
 {\cal H}&=& {\cal H}_0 ~+~{\cal H}_1\nonumber\\
{\cal H}_0 &=& m  \int J_a (\vz) {\d \over {\d J_a (\vz)}} + {2\over \pi} \int _{w,z} 
 {1\over (z-w)^2} {\d \over {\d J_a (\vw)}} {\d \over {\d
J_a (\vz)}} \nonumber\\
&&\hskip .3in+{1\over 2} \int_z : \bdel J^a(z) \bdel J^a(z) :\label{rec1}\\
{\cal H}_1&=&
+ i e \int_{w,z} f_{abc} {J^c(w) \over \pi (z-w)} {\d \over {\d J_a (\vw)}} {\d \over {\d
J_b (\vz)}} \nonumber
\eeqar
The vacuum wave function is taken to be of the form
$\Psi_0 = \exp ({\half} F)$, where
\beqar
F &=& \int f^{(2)}_{a_1 a_2}(x_1, x_2)\ J^{a_1}(x_1) J^{a_2}(x_2) ~+~
\frac{e}{2}\ f^{(3)}_{a_1 a_2 a_3}(x_1, x_2, x_3)\ J^{a_1}(x_1) J^{a_2}(x_2) J^{a_3}(x_3)
\nonumber\\
&&\hskip .2in~+~
\frac{e^2}{4}\ f^{(4)}_{a_1 a_2 a_3 a_4}(x_1, x_2, x_3, x_4)\ J^{a_1}(x_1) J^{a_2}(x_2) J^{a_3}(x_3)
J^{a_4}(x_4)~+~\cdots\label{rec2}
\eeqar
The kernels $f^{(2)}_{a_1 a_2}(x_1, x_2)$, 
$f^{(3)}_{a_1 a_2 a_3}(x_1, x_2, x_3)$, {\it etc}., are posited to have the expansion
\beqar
f^{(2)}_{a_1 a_2}(x_1, x_2) &=& f^{(2)}_{0~a_1 a_2}(x_1, x_2) +
e^2 f^{(2)}_{2~a_1 a_2}(x_1, x_2) +\cdots\nonumber\\
f^{(3)}_{a_1 a_2 a_3}(x_1, x_2, x_3)&=& f^{(3)}_{0~a_1 a_2 a_3}(x_1, x_2, x_3) + e^2 f^{(3)}_{2~a_1 a_2 a_3}(x_1, x_2, x_3)
+\cdots\label{rec3}\\
f^{(4)}_{a_1 a_2 a_3 a_4}(x_1, x_2, x_3, x_4) &=& f^{(4)}_{0~a_1 a_2 a_3 a_4}(x_1, x_2, x_3, x_4) +\cdots\nonumber
\eeqar
We substitue this into the Schr\"odinger equation and, by equating coefficients of terms with equal number of $J$'s, we get a number of recursion relations.
The term corresponding to zero powers of $J$ gives a constant term
which is a normal ordering term for the Hamiltonian.
It is taken account of by the normal ordering indicated in the
potential energy in (\ref{rec1}). The coefficient of the term with one power of
$J$ vanishes by color contractions.
From the term with two powers of the current, we find
\beqar
&& 2m~ f^{(2)}_{a_1 a_2}(x_1, x_2) + 4 \int_{x,y}  f^{(2)}_{a_1 a}(x_1, x) (\bar{\Omega}^0)_{ab}(x,y) f^{(2)}_{b a_2}(y, x_2) +V_{ab} \nonumber \\
&&+e^2 \left[ 6 \int_{x,y} \!\! f^{(4)}_{a_1 a_2 a b }(x_1, x_2, x,y) (\bar{\Omega}^0)_{ab}(x,y) + 3 \int_{x,y} \!\! f^{(3)}_{a_1 a b  }(x_1, x,y) (\bar{\Omega}^1)_{ab a_2}(x,y, x_2)\right] 
= 0\nonumber\\
 \label{rec4}
\eeqar
For $p \ge 3$ the recursion relation is
\beqar
&&m p f^{(p)}_{a_1\cdots a_p} + \sum_{n=2}^{p} n(p+2-n) f^{(n)}_{a_1\cdots a_{n-1} a}(\bar{\Omega}^0)_{ab} f^{(p-n+2)}_{b a_n\cdots a_p} \nonumber \\
&&+ \sum_{n=2}^{p-1} n(p+1-n) f^{(n)}_{a_1\cdots a_{n-1}a} (\bar{\Omega}^1)_{ab a_p} f^{(p-n+1)}_{b a_n\cdots a_{p-1}} \nonumber\\
&&+ e^2 \left[ \frac{(p+1)(p+2)}{2}\ f^{(p+2)}_{a_1\cdots a_p a b}(\bar{\Omega}^0)_{ab} +\frac{p(p+1)}{2}\ f^{(p+1)}_{a_1\cdots a_{p-1} a b} (\bar{\Omega}^1)_{ab a_p}\right] =0 \label{rec5}
\eeqar
In these equations, we have used the abbreviations
\beqar
(\bar{\Omega}^0)_{ab}(x,y) &=& \delta_{ab} \partial_y \bar{G}(x,y) \nonumber \\
(\bar{\Omega}^1)_{abc}(x,y,z) &=& -\frac{i}{2}\ f^{abc} \left[ \delta(z-y) + \delta(z-x)\right] \bar{G}(x,y) \nonumber \\
V_{ab}(x,y) &=& \delta_{ab} \int_z \bar{\partial}_z \delta(z-x) ~\bar{\partial}_z \delta(z-y) \label{rec6}
\eeqar
At the lowest (zeroth) order in $e$, we have to solve (\ref{rec4}) for 
$f^{(2)}_{0\ a_1 a_2}(x_1, x_2) $; this leads to
\footnote{For the holomorphic/antiholomorphic components $k, {\bar k}$, we use, $k = {\half} (k_1 +ik_2)$, ${\bar k} = {\half} (k_1 -i k_2)$. In expressions like
$E_k = \sqrt {k^2 +m^2}$, $k^2$ denotes $k_1^2 + k_2^2$.}

\beq
f^{(2)}_{0\ a_1 a_2}(x_1, x_2)  = \delta_{a_1 a_2}
\left[- {{\bar q}^2 \over m +E_q}\right]_{x_1, x_2}
\label{rec7}
\eeq
This agrees with the kernel used in the Gaussian term in (\ref{sc2a}). Thus, to the lowest order in this expansion, we get the same result for the string tension, namely,
equation (\ref{intro2}).

In this paper we will outline calculations to the next order, i.e., to order $e^2$.
For this we will need the lowest order results for $f^{(3)}$ and $f^{(4)}$.
The recursive solution of equations (\ref{rec4}-\ref{rec6}) to order $e^2$ gives the following lowest order expressions for the cubic and quartic kernels.
\beq
f^{(3)}_{0\ a_1 a_2 a_3}(k_1, k_2, k_3) = -\frac{f^{a_1 a_2 a_3}}{24}\ (2\pi)^2 \delta (k_1+k_2+k_3)\  g^{(3)}(k_1,k_2,k_3)\label{rec50}
\eeq
\beq
f^{(4)}_{0\ a_1 a_2; b_1 b_2}(k_1, k_2; q_1, q_2) = \frac{f^{a_1 a_2 c} f^{b_1 b_2 c}}{64}\ (2\pi)^2 \delta (k_1+k_2+q_1+q_2)\ g^{(4)}(k_1, k_2; q_1, q_2) \label{rec51}
\eeq
where
\beq \label{rec52}
g^{(3)}(k_1,k_2,k_3) = \frac{16}{E_{k_1}\! + E_{k_2}\! + E_{k_3}}\left \{ \frac{\bar k_1 \bar k_2 (\bar k_1 - \bar k_2)}{(m+E_{k_1})(m+E_{k_2})} + {cycl.\ perm.} \right \}
\eeq
and
\begin{equation}\label{rec53}
\begin{array}{cl}
g^{(4)}(k_1, k_2; q_1, q_2)& =\ \vspace{.2in} \displaystyle \frac{1}{E_{k_1}\! + E_{k_2}\! + E_{q_1}\! + E_{q_2}} \\
\vspace{.2in}
&\!\!\!\!\displaystyle \left \{ g^{(3)}(k_1, k_2, -k_1-k_2)\ \frac{k_1 + k_2}{\bar k_1 +\bar k_2}\ g^{(3)}(q_1, q_2, -q_1-q_2) \right . \\
\vspace{.2in}
&\displaystyle -  \left [ \frac{(2\bar k_1 + \bar k_2)\,\bar k_1}{m + E_{k_1}} - \frac{(2\bar k_2 + \bar k_1)\,\bar k_2}{m + E_{k_2}}\right ]\frac{4}{\bar k_1+\bar k_2}\  g^{(3)}(q_1, q_2, -q_1-q_2) \\
&\displaystyle -  \left .   g^{(3)}(k_1, k_2, -k_1-k_2)\ \frac{4}{\bar q_1+\bar q_2}\left [ \frac{(2\bar q_1 + \bar q_2)\,\bar q_1}{m + E_{q_1}} - \frac{(2\bar q_2 + \bar q_1)\,\bar q_2}{m + E_{q_2}}\right ] \right\} \\
\end{array}
\end{equation}
These are defined in terms of the Fourier transforms
\beqar
f^{(3)}_{a_1 a_2 a_3}(x_1, x_2, x_3)&\!\!\!\!\!=\!\!\!\!\!& \int d\mu (k_1,...,k_3) \exp\left( i \sum_i^3 k_i x_i\right) \ f^{(3)}_{a_1 a_2 a_3}(k_1, k_2, k_3)\nonumber\\
f^{(4)}_{a_1 a_2 a_3 a_4}(x_1, x_2, x_3, x_4) &\!\!\!\!\!=\!\!\!\!\!& \int d\mu (k_1,...,k_4) \exp\left( i \sum_i^4 k_i x_i\right)\ f^{(4)}_{a_1 a_2 a_3 a_4}(k_1, k_2, k_3, k_4),\nonumber\\
\label{rec8}
\eeqar
where
\beq
d\mu (k_1, ..., k_n) = {d^2k_1\over (2\pi )^2}\cdots {d^2k_n\over (2\pi )^2}
\label{measure}
\eeq

Once again, the value of $f^{(3)}_{0\ a_1 a_2 a_3}(x_1, x_2, x_3)$ agrees with
previous calculations \cite{KKN2}.
Note also that $f^{(4)}_{a_1 a_2; b_1 b_2}(k_1, k_2; q_1, q_2)$ as defined in (\ref{rec51},\ref{rec53}) is symmertic under independent exchange of the first and second pairs of indices as well
as under the simultaneous exchange $(\{a_1,k_1\}, \{a_2, k_2\}) \leftrightarrow (\{b_1,q_1\}, \{b_2, q_2\})$. We could have certainly made it completely symmetric but
we prefer not to do so to keep the notation simple.

Using the expressions for
 $f^{(3)}_0$, $f^{(4)}_0$ in (\ref{rec4}), the order $e^2$-term in $f^{(2)}$ is
 given by
\beq
f_2^{(2)}(q) = \frac{m}{E_q}\left(\int \frac{d^2 k}{32\pi}\ \frac{1}{\bar k}\ g^{(3)}(q,k,-k-q)\ +\ \int \frac{d^2 k}{64\pi}\ \frac{k}{\bar k}\ g^{(4)}(q,k;-q,-k)\right)\label{rec9}
\eeq
The kernels $f^{(n)}$, $n\geq 5$, become nontrivial only at the next order.

Equations(\ref{rec7}-\ref{rec9}) give our recursive solution for the wave function to order
$e^2$.

The explicit evaluation of (\ref{rec9}) presents no difficulties. Two observations greatly simplify the task. First, the mass $m$ regulates infrared behavior of
the integrals appearing in (\ref{rec9}), thus enabling the expansion in the
external momentum. Second, we are interested only
in the leading ${\cal O}(\bar q^2)$ term. This way we find the following {\it analytic} result
\beq
f^{(2)}(q)\ =\ \frac{\bar q^2}{m}\left(-\frac{63}{32}\ +\frac{25}{4}\,{\rm log}\frac{3}{2}\right)\ +\ {\cal O}(\bar q^2\, q\bar q)\ =\ \frac{\bar q^2}{2 m}\,(1.1308) +\ldots
\label{rec54}
\eeq
Comparing this with the coefficient of ${\cal O}(\bar q^2)$ term in the expansion of the zeroth order kernel we conclude that this is equivalent to a $-113.08\%$
correction. However, other corrections of this order need to be taken into account
before we can reach any conclusion.

\subsection{Computation of expectation values}

In carrying out calculations with $\Psi_0^* \Psi_0 =
e^{{\half}(F+F^*)}$, we have to integrate over the currents $J$ 
(and ${\bar J}$) with the integration measure for the hermitian WZW theory.
This procedure will be worked out shortly.
It will turn out that the expectation values of products involving
only the currents $J$ or only the currents ${\bar J}$ are 
straightforward to evaluate but those involving both $J$'s 
and ${\bar J}$'s are more difficult. For this reason, we shall use a slightly different strategy.

Consider going back to the description in terms of the gauge potentials
$A, \bA$. The ground state wave function is expected to be real, so that the general formula for the expectation value of an observable ${\cal O}$ (which must be
gauge-invariant) can  be written as
\beqar
\la {\cal O}\ra &=& \int d\mu~ \Psi_0^* (A, \bA) {\cal O}(A, \bA) \Psi_0 (A, \bA)\nonumber\\
&=& \int d\mu~ \Psi_0 (A, \bA) {\cal O}(A, \bA) \Psi_0 (A, \bA)
\label{exp1}
\eeqar
Now the integrand is a functional of $A, \bA$, but we can also write it
in terms of $J$ as
\beqar
A &=& M^{\dagger -1} \left[ - {\pi \over c_A} J \right] M^\dagger ~+~
M^{\dagger -1} \del M^\dagger\nonumber\\
\bA &=& M^{\dagger -1} \bdel M^\dagger =  M^{\dagger -1} \left[ ~0~ \right] M^\dagger ~+~
M^{\dagger -1} \bdel M^\dagger
\label{exp2}
\eeqar
$\Psi$ now becomes a function of $M^\dagger$ and $J$. Equation
(\ref{exp2}) shows that we may think of $\{A, \bA\}$ as the (complex)
gauge transform by $M^\dagger$ of $\{ -(\pi /c_A) J, 0\}$.
As argued in \cite{KKN2}, we may then use the gauge invariance of the wave functions to set $M^\dagger$ to $1$, making $\Psi$ a function of $J$.
(In other words, one may think of our transformation to the $J$-variables as a complex gauge choice.)
Once this is done, equation (\ref{exp2}) gives 
\beqar
\la {\cal O} \ra &=& \int d\mu (H) e^{2c_A S_{wzw}(H) }~ \Psi_0(J) \Psi_0(J) ~{\cal O}(J)\nonumber\\
&=& \int d\mu (H) e^{2c_A S_{wzw}(H) }~ e^{F(J)} ~{\cal O}(J)
\label{rec10}
\eeqar
The problem is thus reduced to the computation of the correlators of $J$.
(One may wonder whether and how it is possible to calculate 
with $J$ and ${\bar J}$. It can be done, in principle; some comments about the difficulties of such a procedure are given in the appendix.)

Now, the current $J$ is constrained in terms of the matrix $H$ by the relation
$J = (c_A/\pi ) \del H H^{-1}$; the fields $\vf^a$ in $H= \exp (t^a \vf^a)$
are the unconstrained variables of integration. The direct integration of 
$\Psi_0^* \Psi_0$ over the fields $\vf^a$ is very involved and tedious.
Our strategy will be to rewrite the expectation values in terms of a functional integral over an unconstrained boson field.
We start by rewriting (\ref{rec10}) as
\beqar
\la {\cal O}\ra & =& \displaystyle\left[ {\cal O}({\hat J}) ~ e^{F({\hat J})}\right]~\frac{1}{{\cal Z}}\! \int d\mu (H)\, e^{2c_A S_{wzw}(H) }\, e^{-{c_A\over \pi}
\int {\bar C}^a (\del H H^{-1})^a}\biggr]_{{\bar C} =0}
\label{rec11} \\
{\cal Z} &=&\displaystyle \int d\mu (H)\, e^{2c_A S_{wzw}(H) }\nonumber
\eeqar
where ${\hat J}^a = -\sqrt{2\pi /mc_A} ~{\delta \over \delta {\bar C}^a}$.
The remaining integral can be evaluated as
\beqar
\frac{1}{{\cal Z}}\! \int d\mu (H) e^{2c_A S_{wzw}(H) } e^{-{c_A\over \pi}
\int {\bar C}^a (\del H H^{-1})^a}
&=&\frac{1}{{\cal Z}}\! \int d\mu (H) e^{2c_A S_{wzw}(UH) -2c_A S_{wzw}(U)}
\nonumber\\
&=&\displaystyle e^{-2 c_A S_{wzw}(U)}\label{rec12}
\eeqar
where ${\bar C} = U^{-1}\bdel U$ and
we have used the Polyakov-Wiegmann identity 
\beq
S_{wzw} (H) - {1\over \pi} \int \Tr ({\bar C} \del H H^{-1})
= S_{wzw}(UH) ~-~ S_{wzw} (U)
\label{rec13}
\eeq
The expression $\exp (-2c_A S_{wzw}(U))$ is the inverse of the chiral Dirac determinant in two dimensions. We can therefore represent it in terms of a 
chiral boson field (in two Euclidean dimensions) as
\beq
\exp (-2c_A S_{wzw}(U) ) =
\int [d\vf d{\bar \vf}]~ e^{-\int {\bar \vf} ( \bdel +{\bar C} )\vf}
\label{rec14}
\eeq
The complex boson field $\vf$ transforms as the adjoint representation of $SU(N)$.
Upon using this back in (\ref{rec11}), we find
\beq
\la {\cal O} \ra = \int [d\vf d{\bar \vf}]~  e^{-S(\vf )}~{\cal O}(\sqrt{2\pi /mc_A} ~\bvf t^a \vf )
\label{rec15}
\eeq
where 
\beq
S(\vf ) = \int \bvf \bdel \vf - F(\sqrt{2\pi /mc_A} ~\bvf t^a \vf )
\label{rec16}
\eeq
There is still one correction to be made to this formula. This is because the interactions of the chiral boson, represented by $F$, are such that there is renormalization of the free action $\int \bvf (\bdel +{\bar C} )\vf$.
In such a situation, the use of the integral representation (\ref{rec14})
for the functional integral on the right hand side of 
(\ref{rec11}) is not adequate. At the free level, with no interactions
coming from $e^F$, we can use (\ref{rec14}).
When interactions are added in, we must ask: what representation for the determinant is valid such that we revert to the usual formula (\ref{rec14})
in the kinematic regime where such interactions vanish?
We notice that the $F$ terms come with powers of momenta of the currents and vanish as these go to zero. So we must use a representation for the
determinant which preserves this property.
This means that, instead of (\ref{rec14}), we must use the formula
\beq
\exp (-2c_A S_{wzw}(U) ) =
\int [d\vf d{\bar \vf}]~ \exp\left[ {-\int {\bar \vf} (Z_2 \bdel + Z_1 {\bar C} )\vf}
\right]
\label{rec17}
\eeq
The renormalization constants $Z_1, Z_2$ are anyway equal to $1$
in the absence of the interactions in $e^F$, so there is no contradiction with
the free formula (\ref{rec14}), when there are no interactions. However, 
we can also choose $Z_1, Z_2$ to cancel
any corrections due to loop integrations arising from the interactions $e^F$, so that
we do recover the free formula when the momenta of the currents go to zero.
With this more general formula, we get
\beq
S(\vf ) = \int ~(Z_2 \bvf \bdel \vf + Z_1 \bvf {\bar C}\vf )~-~ F(Z_1\sqrt{2\pi /mc_A} ~\bvf t^a \vf )
\label{rec18}
\eeq
We could set ${\bar C}$ to zero at this point, but we have kept it in 
the above formula to show that we can calculate $Z_1$ from the renormalization of the term $\bvf {\bar C}\vf $, i.e., vertex renormalization, and the same factor should
eliminate all divergences from the interactions $F$.

Equation (\ref{rec18}) is our formula for the action at the chiral boson level.
We can now treat this as a standard two-dimensional field theory and calculate corrections to the various terms. In particular, we are interested in corrections to the
term $F^{(2)}$ which is quadratic in the currents. From the point of view of the
chiral boson, this is a four-point vertex. The term $F^{(2)}$ itself does not have
any powers of $e$, so that, to be consistent, we must take account of all
loops due to this term. This is an additional resummation that has to be done.

As an important example of this resummation, consider the two-point function for the currents, which is given, up to constant factors, by
$\la \bvf t^a \vf (x)~ \bvf t^b \vf (y)\ra$. This may be represented diagrammatically as
\begin{figure}[!t]
\vspace{2pt}
\begin{center}
\includegraphics[height = .2\textwidth, width=.9\textwidth]{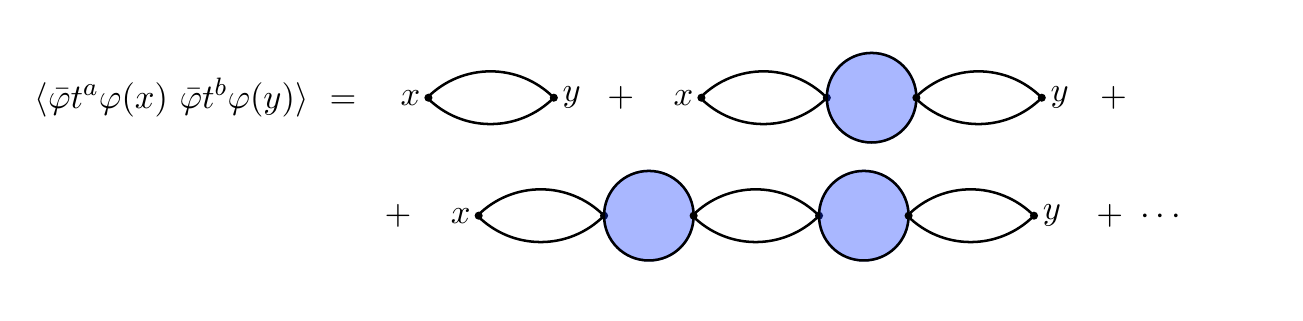}\\
\vskip .1in
\caption{Corrections to the two-point function of currents due to the vertex $F^{(2)}$}
\end{center}
\end{figure}
shown in figure 2, where the solid lines represent $\la \vf ~\bvf\ra$ propagators and the shaded circle is the vertex corresponding to $F^{(2)}$.
The contribution from the free part of the action (\ref{rec18}), represented by the first term
on the right hand side of this equation, 
is
\beqar
\la \bvf t^a \vf (x)~ \bvf t^b \vf (y)\ra &=& \Tr (t^a t^b)  \left( {1\over \bdel }\right)_{xy}
 \left( {1\over \bdel }\right)_{yx}  = - {c_A \over \pi^2} {\delta^{ab}\over (x-y)^2}
 \nonumber\\
 &=& \delta^{ab} {c_A \over \pi} \int {d^2k \over (2\pi )^2} e^{ik(x-y)}~
 {k\over {\bar k}} 
 \label{rec20}
 \eeqar
Using (\ref{rec7}), the contribution of the term with one insertion of the $F^{(2)}$-vertex is
\beq
\la \bvf t^a \vf (x)~ \bvf t^b \vf (y)\ra^{(1)} =
\delta^{ab} {c_A \over \pi} \int {d^2k \over (2\pi )^2} e^{ik(x-y)}~ {k\over {\bar k}} 
~\left[
-{(E_k -m) \over m}\right]
\label{rec21}
\eeq
The summation of the series of terms shown in figure 2 is thus given by
\beq
\la \bvf t^a \vf (x)~ \bvf t^b \vf (y)\ra = 
\delta^{ab} {c_A \over \pi} \int {d^2k \over (2\pi )^2} e^{ik(x-y)}~
 {k\over {\bar k}}~ \left( {m\over E_k}\right)
 \label{rec22}
 \eeq
The presence of the $m/E_k$ factor improves ultraviolet convergence of integrals and it will also suppress the numerical values of various corrections. 
In any diagram, the vertex corresponding to a current $\bvf t^a \vf$ will have such a factor of $m/E_k$.
This follows from noticing that the current in any diagram has a series of terms (due to insertions of $F^{(2)}$) correcting it, as shown in figure 3.
The summation of these terms gives the result

\beqar
\bvf t^a \vf (x)\Bigr]_{eff} &=& \bvf t^a \vf (x) ~- ~ \int {d^2k\over (2\pi )^2 }
 e^{ik(x-z)} ~{E_k -m \over m}  ~(\bvf t^a \vf )(z) ~+~\cdots\nonumber\\
 &=& \int {d^2k\over (2\pi )^2 }
 e^{ik(x-z)} \left[ 1- {E_k -m \over m}+ \left({E_k - m\over m}\right)^2 
 +\cdots\right]  (\bvf t^a \vf )(z)\nonumber\\
 &=& \int {d^2k\over (2\pi )^2} 
 e^{ik(x-z)} ~{m\over E_k} ~ (\bvf t^a \vf )(z)
 \label{rec23}
 \eeqar
For the two-point function for the currents, we must use the corrected current
given by this equation only at one vertex; otherwise, there will be double-counting.
This is similar to the case of Schwinger-Dyson equations in, say, electrodynamics, where the vertex corrections to the vacuum polarization only apply at one vertex.
\vskip .1in
\begin{figure}[!h]
\begin{center}
\includegraphics[height = .2\textwidth, width=.87\textwidth]{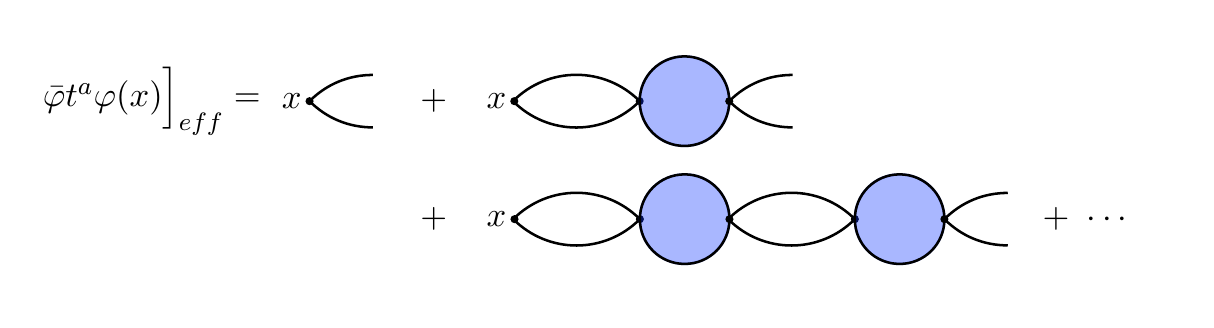}\\
\vskip .1in
\caption{The effective current vertex}
\end{center}
\end{figure}
\subsection{The renormalization terms}
We shall now consider briefly the renormalization constants
$Z_1$, $Z_2$.

The self-energy and vertex corrections for the field $\vf$ may be represented by the graphs shown in
figure 4. Direct computation then shows that the 
corrected action, to this order, may be written as
\beqar
S(\vf )&=& \int Z_2 \bvf \bdel \vf  \left[ 1- {4\pi c_R\over mc_A} \int {d^2k\over (2\pi)^2}
{1\over E_k +m} {m \over E_k}\right]\nonumber\\
&&\hskip .2in+ \int Z_1 \bvf {\bar C}\vf \left[ 1- {4\pi \over mc_A} (c_R -{\half} c_A) \int {d^2k\over (2\pi)^2}
{1\over E_k +m} {m \over E_k}\right]~+\cdots
\label{rec24}
\eeqar
For the moment, we have taken the fields $\vf$ to be in an arbitrary representation $R$. Notice that, if we consider only the terms
proportional to $c_R$, we have $Z_1= Z_2$, as in electrodynamics.
However, because of the $c_A$-terms, $Z_1$ and $Z_2$ end up as different constants. Such  difference between the wave function renormalization and the vertex renormalization is well known in nonabelian theories, with exactly the same mismatch as we find here. Reverting now to the adjoint representation for the fields,
we can identify
\beqar
Z_1&\approx& 1 + {2\pi \over m} \int {d^2k\over (2\pi)^2}
{1\over E_k +m} {m \over E_k}\nonumber\\
Z_2&\approx& 1 + {4\pi \over m} \int {d^2k\over (2\pi)^2}
{1\over E_k +m} {m \over E_k}
\label{rec25}
\eeqar
The potential divergences are logarithmic; if $Z_1$, $Z_2$ are chosen to cancel them, as above, then there are no further divergences. In fact, the corrections to the
vertices $F^{(n)}$, which is what we are interested in, are finite.

\begin{figure}[!t]
\begin{center}
\includegraphics[height = .5\textwidth, width=.82\textwidth]{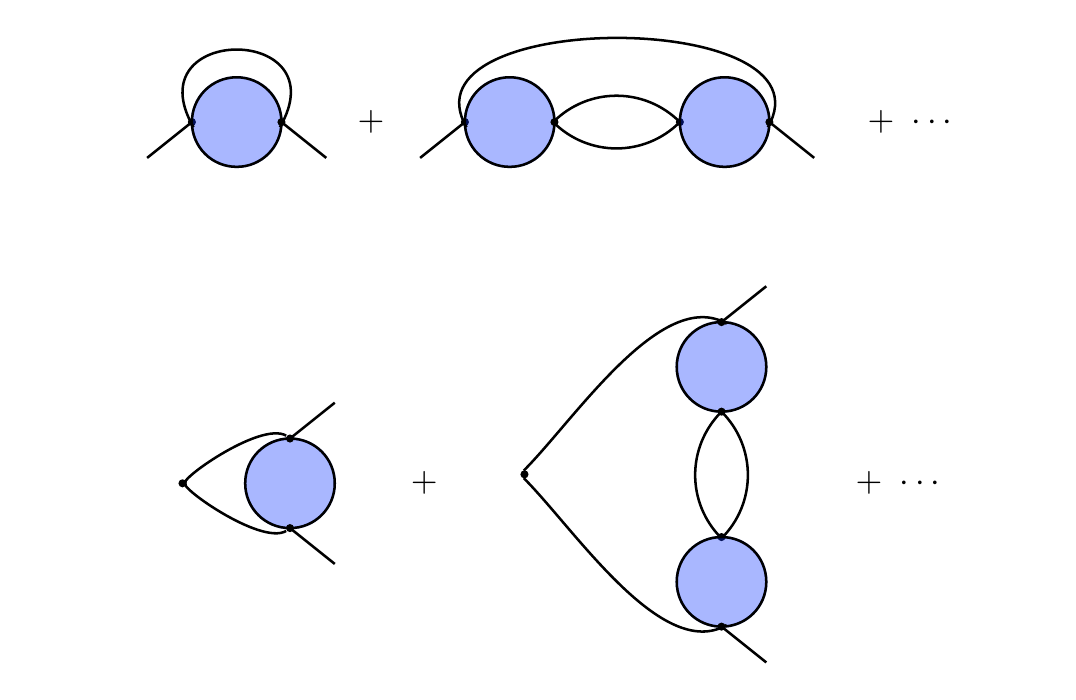}\\
\vskip .1in
\caption{Self energy (upper) and vertex (lower) corrections for the field $\vf$}
\end{center}
\end{figure}

\subsection{Evaluating loop corrections}

We have outlined a procedure for carrying out the summation of terms arising from insertions of the vertex $F^{(2)}$, which has to be done to all orders.
The self-energy and vertex corrections of the type shown in figure 4 are compensated by the choice of the renormalization constants. After
setting ${\bar C}=0$, there are still factors of
$Z_1$ in the vertices $F$. These are compensated by the corresponding insertions of $F^{(2)}$ of the $t$-channel type or vertex type. Likewise, there will be self-energy insertions on the propagator lines which are compensated by the $Z_2$ factors.
The effect of the insertions of $F^{(2)}$ is then reduced to terms of the type
in figures 2 and 3. These can be taken care of by factors of $m/E_k$.
Thus, we have a fairly simple procedure for calculating loop corrections.
We write down the loop corrections due to the $F^{(3)}$, $F^{(4)}$ vertices, etc.
Then for the current vertices, we insert factors of $m/E_k$ appropriately;
each current vertex has such a factor, where $k$ is the difference of the momenta of the $\vf$ and $\bvf$ legs of the vertex. This counting applies in all cases, except
when the two-point function of currents arises, in which case there is only one such factor.
The expressions for the loop corrections, so modified, include the effects of arbitrary number
of $F^{(2)}$ insertions and can be evaluated to give the corrections to various physical quantities.

\section{Corrections to the string tension}

We now turn to the corrections to the string tension which arise from corrections of the first type, i.e., from propagator corrections, as shown in figure 1.
For this purpose, we must calculate, in the $\vf$-language, the corrections to
$F^{(2)}$. Actually, for the string tension, the low momentum limit of such corrections are adequate. The relevant terms can be easily
identified by the Feynman diagrams. To order $e^2$, only the $F^{(3)}$ and $F^{(4)}$ vertices are relevant, in addition to $F^{(2)}$ of course.
The basic loop diagrams are as shown in figures 5-9, where they are arranged by powers of $m/E_k$.
The action $S$ for the calculations is given by 
equation (\ref{rec18}).

We shall now give the contributions of each diagram to the two-point kernel
$f^{(2)}_{a_1 a_2}(x_1, x_2)   = \delta_{a_1 a_2} [f^{(2)}(q)]_{x_1,x_2}$.
Actually, we are interested in the corrections to the string tension. For this purpose,it is sufficient to consider contributions to the low-momentum limit of
$f^{(2)}_0(q)$, which is given by
\beq
f^{(2)}_0(q) = 
-\left[ {{\bar q}^2 \over m +E_q}\right]
\approx - {{\bar q}^2 \over 2m}
\label{corr1}
\eeq
Since the corrections generated by the loop diagrams are quite complicated algebraic expressions, we will give a summary of the low-momentum limits here. Details of the calculations will be presented in an appendix.

First of all, notice that the first correction to $f^{(2)}$ from the recursive solution,
as obtained in (\ref{rec54}), gives
\beq
\Delta f^{(2)}_{rec} \approx {{\bar q}^2 \over 2m} (1.13082) + \cdots
\label{corr2}
\eeq
The contribution from diagram 1 is given by
\beq
{\rm Diagram ~1} \approx {{\bar q}^2 \over 2m} (-0.58118) + \cdots
\label{corr3}
\eeq
This comes from an integral with one power of $m/E_k$.
\begin{figure}[!b]
\begin{center}
\includegraphics[height = .2\textwidth, width=.48\textwidth]{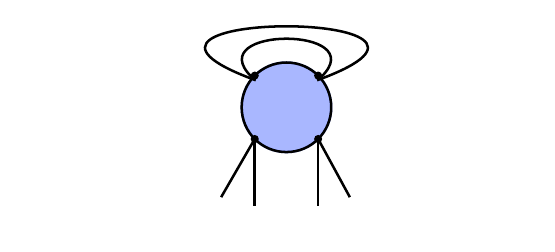}\\
\end{center}
\caption{Diagram $1$ which is the contribution
with one factor of $m/E_k$.}
\label{Diag1}
\end{figure}

Diagrams 2a and 2b, shown in figure 6, contain two powers of $m/E_k$.
The contribution from these diagrams is given as
\beqar
{\rm Diagram~ 2a} &\approx& {{\bar q}^2 \over 2m} (-0.47835) + \cdots
\nonumber\\
{\rm Diagram ~2b}&\approx& {{\bar q}^2 \over 2m} (0.20169) + \cdots
\label{corr4}
\eeqar

\begin{figure}[!t]
\begin{center}
\includegraphics[height = .2\textwidth, width=.84\textwidth]{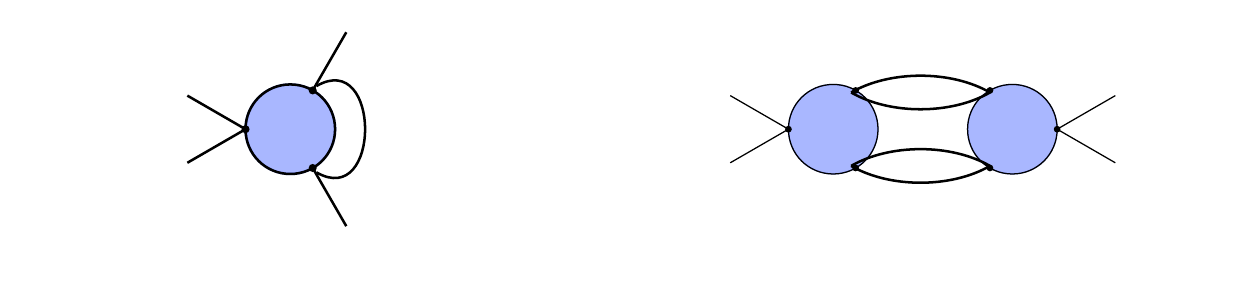}\\
\end{center}
\caption{Diagrams 2a and 2b which are contributions with two
powers of $m/E_k$.}
\label{Diag2ab}
\end{figure}
Notice that diagram 2a has the current ${\bar \vf} t^a \vf$ at one vertex, while from the other two
vertices we get nonlocal expressions like ${\bar \vf}(x) t^a \vf (y)$. Thus we will need to expand this in terms of the currents to identify the contribution (\ref{corr4}). We have carried out this expansion; it is essentially an operator product expansion and will be discussed in appendix B.

There is one term, namely, diagram 3, which has integrals with
$(m/E_k)^3$. Here too we will need an expansion of the product ${\bar \vf }(x) t^a\vf (y)$ in terms of the currents to pick up contributions of the type
$\bdel J \bdel J$. Once this is done, the contribution of this diagram is seen to be
\beq
{\rm Diagram ~3} \approx {{\bar q}^2 \over 2m} (-0.23569) + \cdots
\eeq

\begin{figure}[!b]
\begin{center}
\includegraphics[height = .15\textwidth, width=.36\textwidth]{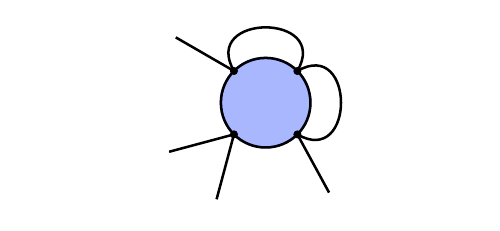}\\
\end{center}
\caption{Diagram $3$ which is the contribution with three powers of
$m/E_k$.}
\label{Diag3}
\end{figure}

\begin{figure}[!t]
\begin{center}
\includegraphics[height = .4\textwidth, width=.85\textwidth]{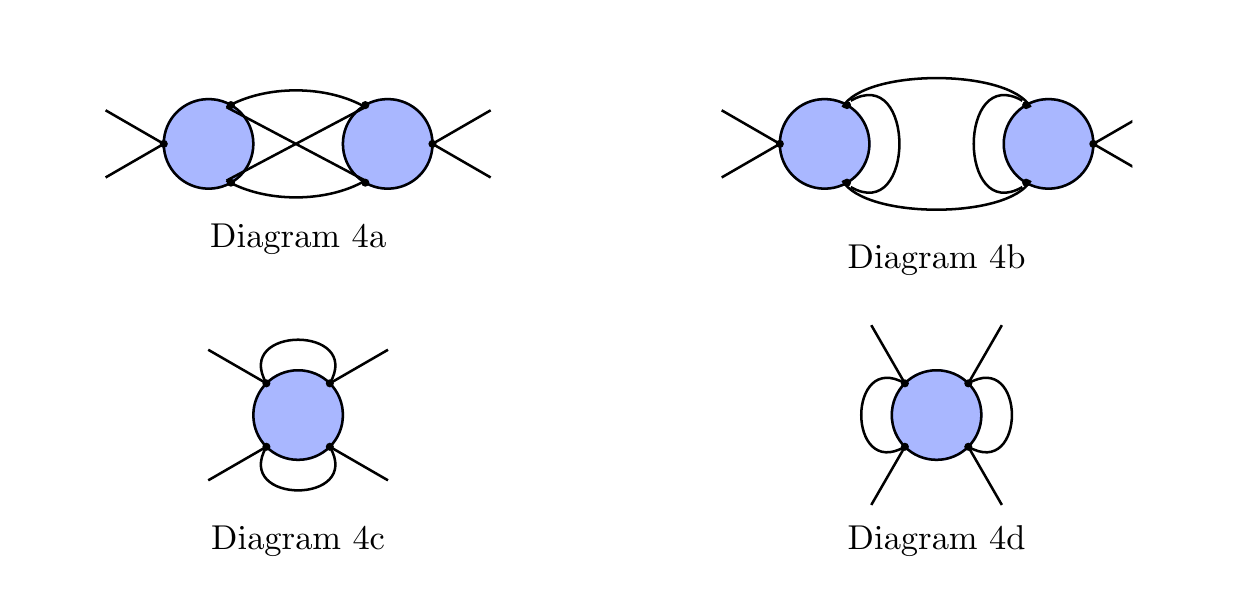}\\
\caption{Diagrams with four powers of $m/E_k$ in the integrals.}
\label{Diag4abcd}
\end{center}
\end{figure}

\begin{figure}[!b]
\begin{center}
\includegraphics[height = .2\textwidth, width=.8\textwidth]{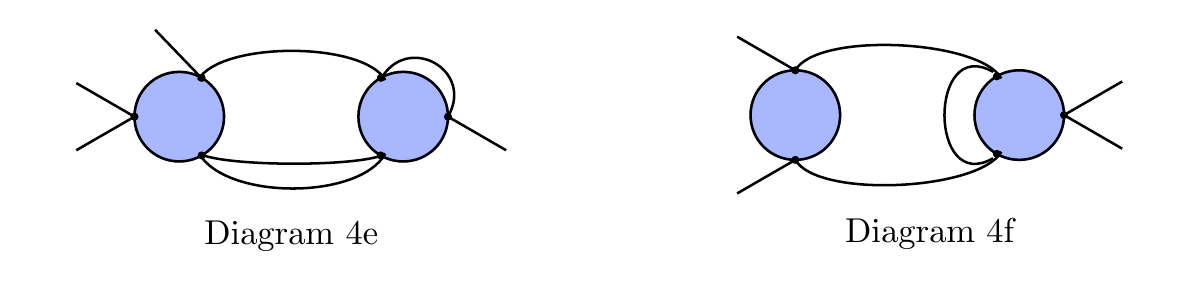}\\
\caption{Two more diagrams with four powers of $m/E_k$ in the integrals.}
\label{Diag4e}
\end{center}
\end{figure}

There are six diagrams at the level of the $(m/E_k)^4$, as shown in figures 8 and 9.
Of these, diagram 4a is zero by the structure of
color contractions. The contributions from the next four diagrams are as follows.
\beq\label{corr6}
\begin{split}
{\rm Diagram ~4b} &\approx  {{\bar q}^2 \over 2m} (0.02083) + \cdots\\
{\rm Diagram ~4c} &\approx  {{\bar q}^2 \over 2m} (-0.06893) + \cdots\\
{\rm Diagram ~4d}&\approx  {{\bar q}^2 \over 2m} (-0.01216) + \cdots\\
{\rm Diagram ~4e} &\approx {{\bar q}^2 \over 2m} (0.06824) +\cdots\\
\end{split}
\eeq
Diagram 4f has some subtleties and we shall take up its calculation shortly.

At this point, it is useful to take stock of where we are in terms of these corrections.
It is instructive to look at the result in terms of partial sums of contributions to
$f^{(2)}$ to a given order in powers of $m/E_k$. Let $\Delta f^{(2)}_n$ denote
the sum of contributions including up to $n$ factors of $m/E_k$ in the integrand.
We then find $\Delta f^{(2)}_n = ({\bar q}^2/2m) C_n$, with
\beq
\begin{split}
C_0 &=  1.13082\\
C_1&= 0.54964\\
C_2&= 0.27298\\
C_3&= 0.03729\\
\end{split}
\label{corr7}
\eeq
Notice that the corrections are systematically getting smaller. 
The first term $C_0$ seems alarmingly large, but our main point is that it should not be
considered in isolation. In fact, it is easy to check that it arises
from the cubic term in ${\cal H}_1$ which corresponds to
the term $f^{abc} J^c (x) /(x-y)$ in the operator product expansion
of $J^a (x) J^b (y)$. Diagram 2a, for example, has a term which is exactly
of this type; if we neglect the factors of $m/E_k$, it would cancel
the part of
$C_0$ arising from $g^{(3)}$ exactly. This shows that various contributions should be 
considered within certain natural groupings.
Our expansion does not put them together in such a way from the beginning, 
but the partial sums
are essentially doing this.
(Individual diagrams give seemingly large values. This is in consonance
with \cite{greensite}; the form of the wave function used there, we suspect,
is picking out only the contribution due to some of the diagrams we have.)

The corrected value of the string tension is
\beq
\begin{split}
\sqrt{\sigma_R} &=  e^2 \sqrt{ c_A c_R\over
4\pi }~~ \bigl( 1 + 0.173 +\cdots \bigr) \hskip .3in {\rm for}~ \Delta f^{(2)}_2\\
&=  e^2 \sqrt{ c_A c_R\over
4\pi }~~ \bigl( 1 +0.019 +\cdots \bigr) \hskip .3in {\rm for}~ \Delta f^{(2)}_3\\
\end{split}
\label{corr8} 
\eeq
We see that the numerical value of $\sqrt{\sigma_R}$, up to the diagrams of order
$(m/E_k)^3$, has approximately
a $2\%$ correction compared to (\ref{intro2}); this correction, however, moves it further away from
the recent lattice estimates.
Of course, we still have to include
the diagrams with $4$ factors of $m/E_k$. However, the main point we want to
make at this point is that the corrections are small, of the order of a few percent,
and that there is a systematic expansion scheme which gives a sensible ordering of the
various corrections.

The total contribution from the diagrams 4a to 4e is
\beq
\Delta f^{(2)}_{4a-4e} = {{\bar q}^2 \over 2m} ( 0.00798)
\label{corr9}
\eeq
This is a numerically small correction. 

As for diagram 4f, notice that it can 
be thought of as arising from an effective vertex represented by diagram 2a 
which is then Wick contracted with the
$F^{(2)}$ vertex. Among the diagrams we are considering, this is the only one which involves
contraction with an $F^{(2)}$ vertex. This is why it is special.
Now, imagine that we are considering the low momentum limit of
diagram 2a, namely, $\Delta f^{(2)}_{2a}$ given in equation
(\ref{corr4}). When further calculations are done with the
corrected $f^{(2)}$ of the form $f^{(2)} + \Delta f^{(2)}_{2a}$, we naturally
encounter Wick contractions between the fields which go with the
two terms. This generates a term which is essentially the same as the 
diagram 4f, but with the loop integration over the new loop momenta restricted to
low momentum values.
 (We are first taking the low momentum limit and then 
calculating the subsequent
diagram, so the loop integrals in this latter calculation have to be cut off at some value.
This is why only the low momentum part of the integrations occur.)
In other words, a part of diagram 4f will be generated in subsequent calculations we
do in using the effective action for low momentum correlators.
Thus, to avoid double-counting, only the contribution from the high momentum part of the 
integration over the second loop momentum
in diagram 4f should be included as a correction at this stage. Unfortunately, the
separation between high and low momentum in the loops is somewhat ambiguous. 
If we calculate diagram 4f with the integrations done only over momenta
$> m$, the value is 
\beq
\Delta f^{(2)}_{4f} \approx {{\bar q}^2 \over 2m} ( - 0.1037)
\label{corr10}
\eeq
If this value is used, we find
\beq
\Delta f^{(2)}_{4} =  {{\bar q}^2 \over 2m}  C_4 \approx {{\bar q}^2 \over 2m} (-.05843)
\label{corr11}
\eeq
The corrected string tension is then given by
\beq
\sqrt{\sigma_R} =  e^2 \sqrt{ c_A c_R\over
4\pi }~~ \bigl( 1 - 0.02799 +\cdots \bigr) \hskip .3in {\rm for}~ \Delta f^{(2)}_4
\label{corr12}
\eeq
This is a $2.8\%$ correction with the correct sign, moving it closer to the lattice value.
We are actually overshooting the lattice value by some amount, but it should be kept in mind that
the estimate of the last correction in (\ref{corr10}) has some ambiguity due to
the choice of the point at which we should cut off the low momentum integrations.
(For the sake of the argument,
if $2m$ is used as the cut off, the value of $\sqrt{\sigma_R}$ differs from
(\ref{intro2}) by $-0.00290$, about one-third of one percent.)
Clearly it is important to  settle the issue of this cut-off on {\it a priori} grounds.
However, one can see that this issue will not affect the results by more than a few percent.
An estimate of how bad this ambiguity could be is given by calculating the integral 
over all values of the loop momentum. The value of $\Delta f^{(2)}_{4f}$ is then 
approximately $({\bar q}^2 / 2m)\times ( - 0.166)$. This could decrease the value of
$\sqrt{\sigma_R}$ by another $3\%$. 
In other words, there is a certain stability to the analysis.
The ambiguity of where the low momentum cut off should be does not change the value
by more than a few percent.

Can we get the corrected string tension to match the lattice value exactly?
First of all, for this, it is necessary to
analyze the question of the low momentum cut off in the integral in
diagram 4f
more carefully. There are also other corrections such as a number of diagrams with $5$ and higher number of  factors
of $m/E_k$. There are also corrections of order $e^4$; these are expected to remain small
because diagrams at this order also have more factors of $m/E_k$.
But we are talking about corrections at the one or two percent level and to get exact agreement
at this level of precision requires consideration of many such contributions.

\section{Summary and Discussion}

The analysis of the
Hamiltonian formulation for Yang-Mills theory in $2+1$ dimensions
carried out here has helped to
place the computation of the wave function and the string tension in the context of a systematic expansion.
This expansion involves a splitting of the Hamiltonian into
${\cal H}_0$, which is quadratic in the current or functional derivatives with respect to it,
and ${\cal H}_1$, which is cubic in these quantities.
The Schr\"odinger equation is then solved recursively in powers of the coupling constant
$e$ and the mass parameter
$m$; these two quantities are treated as independent at this stage, the relation
$m = e^2 c_A/2\pi$ is used only at the end.
This helps to systematize the expansion parametrically.
We have expressed the evaluation of expectation values with the vacuum wave function 
as the functional integral of a two-dimensional chiral boson theory. 
The calculation of low momentum correlators can then be visualized  as a two step process.
First, we define
an effective action which incorporates the loop corrections in the usual way;
this action can be used for calculating low momentum correlators as a second step.

One of the interaction vertices, namely the term which is quadratic in the currents,
has to be included to all orders in most of the calculations. The effect of this is to introduce
additional factors of $m/E_k$ (for an appropriate momentum $k$)
in the loop integrals. This improves convergence of the integrals and makes their
numerical values
smaller. As a result, it is useful to classify
diagrams further by the number of such factors.
We have calculated the first set of corrections in such a scheme, to first order in
the recursive solution for $\Psi_0$ (i.e., to first order in $e^2$),
including all terms with up to $4$ factors of $m/E_k$
in the
loop corrections which contribute to order $e^2$.
The result is a correction to the string tension which is approximately
in the range $-0.3\%$ to $-2.8\%$ in $\sqrt{\sigma_R}$, compared to the value given in
(\ref{intro2}). 
This makes the result consistent with the recent lattice
estimates.

Regarding theoretical uncertainty, we note that 
there is an ambiguity in
one of the integrals, as to precisely where the separation between
what is identified as ``low momentum" and what is identified as
``high momentum" is to be imposed.
This needs to be clarified, even though we see that the variance 
due to this can be no more than a percent or two. 
 There are also corrections involving $5$ or more powers of
$m/E_k$.
There are also higher terms (of order $e^4$) in the recursion 
and corresponding loop corrections.
All these corrections are expected to be small, based on the fact that there are more powers of
$m/E_k$, but when talking about corrections of the order of a percent, they can have
a significant effect.

The true value of our analysis is to show that corrections can be systematized and remain under
control and are numerically small. In this sense, we have demonstrated the feasibility of systematic calculations in our Hamiltonian approach to Yang-Mills theory. The final value of the string tension, to the order we have calculated, is also consistent with the recent lattice estimates.

This work was supported in part
by the National Science Foundation
grants PHY-0555620 and PHY-0758008 and by PSC-CUNY grants.

\section*{\normalsize APPENDIX A}

Our calculations have all been done in terms of correlators of $J$.
Once the wave function has been calculated in terms of
$J$, $\Psi^*$ naturally involves ${\bar J}$ and one may ask whether it is possible to do calculations using $\Psi^* \Psi$.
In principle, this is definitely possible. Since the complete
wave function $\Psi_0 (J)$ is expected to be holomorphically invariant, we can re-express $\Psi_0^* ({\bar J})$ back in terms of $J$, and we only have to evaluate correlators of $J$.
The difficulty, however, is that the separation of the Hamiltonian into ${\cal H}_0$
and ${\cal H}_1$ is not holomorphically invariant and since we are only calculating $\Psi_0$ to a certain order in $e^2$, we do have to consider the issue of 
expressing ${\bar J}$ in terms of $J$. 
(The fact that this separation is not consistent with holomorphic invariance is not
a problem in itself; it preserves the invariance in an order-by-order fashion.
Recall that a similar separation is standard in gauge theories whereby one loses the full gauge invariance at a particular order.
However, gauge invariance is preserved, in a consistent fashion; i.e.,
up to terms of higher order. The Ward identities, likewise, connect vertices
and Green's functions of different orders. A similar situation is expected in our case.)

Now, from the definition of the currents,
$\del {\bar J}^a = \bdel J^b H^{ba}$, where $H^{ba} = 2 \Tr (t^b H t^a H^{-1})$
is the field $H$ as an adjoint matrix.
One way to proceed would be to rewrite ${\bar J}^a$ in terms of
$J^a$ as ${\bar J}^a = {1\over \del}  ( \bdel J^b H^{ba})$ and further express
$H$ in terms of  $J$ as an infinite series.
$\Psi_0^* \Psi_0$ then becomes a function of $J$'s and expectation values can be evaluated.
The difficulty with this approach is that one may need to take account of the full infinite series to get meaningful results.
As an example, recall the calculation of the two-point function of the 
group-valued field $g$ in a unitary WZW model.
We can write $g= \exp (it^a \theta^a)$ and expand in powers of
$\theta^a$ and calculate the correlators.
To the lowest order this gives a logarithmic function,
$\la g(x) g(y) \ra \sim \la \theta^a (x) \theta^b (y) \ra \sim \delta^{ab} \log (x-y)$.
However, the whole series can be summed up to give
a power law result which incorporates the anomalous dimension of
$g$ correctly. (The summation is usually done by more efficient means
such as the Polyakov-Wiegmann identity or the Knizhnik-Zamolodchikov
equation, but that is a separate issue.)
In the case of our field $H$, we expect a similar situation.
We have checked that at least some of the lower order terms
lead to logarithms which are likely to be spurious.
The full resummation is, however, beyond what can be done at present.
Therefore, the method used in subsection 3.2 is significantly
better.

If calculations are done, taking into account the relationship 
between $J$ and ${\bar J}$ in full, then we expect that the 
two ways of calculating, namely, using $\Psi_0^* \Psi_0$ with ${\bar J}$
transformed back to $J$'s as above, or
writing the expectation value in terms of $J$ by setting
$M^\dagger $ to $1$ as done in subsection 3.2, will yield the same result.

\section*{\normalsize APPENDIX B}
\def\theequation{B\arabic{equation}}
\setcounter{equation}{0}

In this appendix we present details of the computation of the diagrams shown in Figures \ref{Diag1}-\ref{Diag4abcd}. We start by collecting various expressions relevant for this discussion.

We consider the two-dimensional chiral boson theory (\ref{rec16}) with interaction term
\beq
F(\bar\vf T^a\vf)\ =\ F^{(2)}(\bar\vf T^a\vf) +\frac{e}{2}\: F^{(3)}(\bar\vf T^a\vf) +\frac{e^2}{4}\: F^{(4)}(\bar\vf T^a\vf) +\ \ldots
\label{B1}
\eeq
where interaction vertices $F^{(2)},  F^{(3)},  F^{(4)}$ are given by
\beqar\label{BF2}
F^{(2)}\!\!\!\! &=&\!\! \frac{2\pi}{m c_A} \int d\mu (q)\ (\bar\vf T^a\vf)_q\: f^{({2})}_0(q)\: (\bar\vf T^a\vf)_{-q}\\\label{BF3}
F^{(3)}\!\!\!\! &=&\!\! \left(\frac{2\pi}{m c_A}\right)^{3/2} \int d\mu (k_1, k_2, k_3)\ (\bar\vf T^{a_1}\vf)_{-k_1} (\bar\vf T^{a_2}\vf)_{-k_2}(\bar\vf T^{a_3}\vf)_{-k_3}\nonumber\\
&&\hspace{2in} \times f^{(3)}_{0\ a_1 a_2 a_3} (k_1,k_2,k_3)\\\label{BF4}
F^{(4)}\!\!\!\! &=&\!\! \left(\frac{2\pi}{m c_A}\right)^2\! \int d\mu (k_1, k_2, q_1, q_2)\ (\bar\vf T^{a_1}\vf)_{-k_1} (\bar\vf T^{a_2}\vf)_{-k_2} (\bar\vf T^{b_1}\vf)_{-q_1}(\bar\vf T^{b_2}\vf)_{-q_2} \\
&& \hspace{2in} \times\ f^{(4)}_{0\ a_1 a_2 b_1 b_2} (k_1,k_2;q_1,q_2) \nonumber
\eeqar
with kernels $f^{(2)},  f^{(3)},  f^{(4)}$ as in eqs. (\ref{rec7}-\ref{rec53}), and
\beq
d\mu (q) = \frac{d^2 q}{(2\pi)^2}, \hskip .3in d\mu (k_1, k_2) =
\frac{d^2 k_1}{(2\pi)^2} \frac{d^2 k_2}{(2\pi)^2}, \hskip .3in {\rm etc.}
\label{B2}
\eeq
For the Wick contraction of the chiral boson fields and currents we have
\beqar
\langle \bar\vf^a(r)\ \vf^b(s)\rangle &=& \delta^{a b}\, (2\pi)^2 \delta (r+s)\: \frac{1}{i\,\bar r}\nonumber\\
\langle (\bar\vf T^{a}\vf)_{k}\ (\bar\vf T^{b}\vf)_{p}\rangle &=& \delta^{a b}\: (2\pi)^2 \delta (k+p)\ \frac{c_A}{\pi}\frac{k}{\bar k}\, \frac{m}{E_k}  
\label{B3}
\eeqar 

\subsection*{Diagram 1}

Diagram $1$ in Figure \ref{Diag1} is quite easy to evaluate. Contraction of one pair of bosonic currents in $F^{(4)}(\bar\vf T\vf)$ leads to the following expression
\beq\label{D11}
{\rm Diagram\ 1} = \frac{2\pi}{m c_A} \int d\mu (q)\ (\bar\vf T^a\vf)_{-q}\, (\bar\vf T^a\vf)_q \ \left\{\int \frac{d^2 k}{64\pi} \ \frac{k}{\bar k}\: \frac{m}{E_k}\ g^{(4)}(q,k; -q,-k)\right\}
\eeq
The only less trivial step in deriving (\ref{D11}) is that combinatorial factor is $\frac{4 \times 2}{2} =4$ instead of the na\"ive expectation $\frac{4 \times 3}{2} =6$. This is so because we chose not to symmetrize completely the quartic kernel $f^{(4)}_{0\ a_1 a_2 b_1 b_2}(k_1, k_2; q_1, q_2)$ . 

Comparing (\ref{D11}) with $F^{(2)}(\bar\vf T\vf)$ from eq.(\ref{BF2}) we immediately conclude that correction to quadratic kernel $f^{(2)}(q)$
due to this diagram is 
\beq\label{D12}
\Delta f^{(2)}(q) = \int \frac{d^2 k}{64\pi}\ \frac{k}{\bar k}\: \frac{m}{E_k}\ g^{(4)}(q,k; -q,-k)
\eeq
The leading, ${\cal O}(\bar q^2)$, term in the low-momentum expansion of (\ref{D12}) can be computed {\it analytically}; we find
\beq
\Delta f^{(2)}(q) = \frac{\bar q^2}{2 m}\left(\frac{81}{16}+ \frac{17}{2}\ {\rm log}\, 2 -\frac{21}{2}\ {\rm log}\, 3\right)\ +\ {\cal O}(\bar q^2\ q\bar q)\ =\ \frac{\bar q^2}{2 m} (-0.581178)\ +\ \ldots
\eeq

\subsection*{Diagram 2b}

Of the two diagrams with two factors of $m/E_k$ shown in Figure \ref{Diag2ab} , 
diagram $2b$ is straightforward, so we will evaluate it first.
 The Wick contraction of two pairs of bosonic currents gives 
\beqar
{\rm Diagram\ 2b} &=& \frac{2\pi}{m c_A} \int d\mu (q)\ (\bar\vf T^a\vf)_{-q}\, (\bar\vf T^a\vf)_q \nonumber\\
&&\hspace{.5in} \times\left\{-\int \frac{d^2 k}{128\pi}\ \frac{k(k+q)}{\bar k(\bar k +\bar q)}\ \frac{m}{E_k\, E_{q+k}} \left[g^{(3)}(q,k,-q-k)\right]^2\right\}
\label{B4}
\eeqar
Again, comparing this expression with (\ref{BF2}) we conclude that correction to $f^{(2)}(q)$ due to Diagram $2b$ is 
\beq\label{D2b2}
\Delta f^{(2)}(q) =\ -\int \frac{d^2 k}{128\pi}\ \frac{k(k+q)}{\bar k(\bar k +\bar q)}\ \frac{m}{E_k\, E_{q+k}} \left[g^{(3)}(q,k,-q-k)\right]^2
\eeq
The leading, ${\cal O}(\bar q^2)$, term in the expansion of (\ref{D2b2}) can be calculated {\it analytically} as
\beq
\Delta f^{(2)}(q) = \frac{\bar q^2}{2 m}\left( -\frac{23}{8} + \frac{13}{4}\: {\rm log}\, 2 + \frac{3}{4}\: {\rm log}\, 3\right) + \ldots = \frac{\bar q^2}{2 m}\,(0.201688) +\ldots
\label{B5}
\eeq

\subsection*{Diagram 2a}

Diagram $2a$ provides the first nontrivial example of diagrams we have to deal with. First of all, notice that it is based on a single 
cubic vertex $F^{(3)}$, and, as such, this diagram does not have a direct analogue in standard perturbative QCD. Second, while it is elementary to 
write an explicit expression which corresponds to this diagram, i.e.,
\beq\label{D2a1}
{\rm Diagram\ 2a} = \frac{2\pi}{m c_A} \int d\mu (q,r) \ (\bar\vf T^a\vf)_{-q}[\bar\vf(r) T^a\vf(-r+q)+ \bar\vf(-r+q) T^a\vf(r)]\: \Pi(r,q)
\eeq
with
\beq\label{D2a2}
\Pi(r,q) = \int \frac{d^2 k}{64\pi}\ \frac{g^{(3)}(k,q,-k-q)}{\bar r + \bar k}\ \frac{m^2}{E_k\, E_{k+q}} ,
\eeq
the interpretation of this expression requires some care. The problem is that the term which is bilinear in $\bar \vf$ and $\vf$, namely,
\beq\label{D2a3}
{\cal I} \equiv \int \frac{d^2 r}{(2\pi)^2}\ [\bar\vf(r) T^a\vf(-r+q)+ \bar\vf(-r+q) T^a\vf(r)]\: \Pi(r,q),
\eeq 
is not simply proportional
to a current $(\bar \vf T^a \vf)_q$ and therefore (\ref{D2a1}) cannot be directly compared to (\ref{BF2}). This is because $\Pi(r,q)$ depends on $r$ in a nontrivial way
 \footnote{Imagine for a moment that we may forget about the $r$-dependence 
 of $\Pi(r,q)$ and take, for example,
 $\Pi(0,q)$ instead. In this case (\ref{D2a3}) nicely factorizes into $2\,\Pi(0,q)\: (\bar\vf T^a\vf)_{q}$, and we see that (\ref{D2a3}) would be proportional
to a current $(\bar\vf T^a\vf)_{q}$.}.
We note that $\Pi(r,q)$
admits a power series expansion in $r$, which means that, in addition to the current $(\bar \vf T^a \vf)_q$, the expansion of (\ref{D2a3}) also contains other local operators of the type
$({\bar\partial}^n{\partial}^m \bar \vf T^a \vf)_q$ and $(\bar \vf T^a {\bar\partial}^n{\partial}^m \vf)_q$. 
But, in our two-dimensional theory,  any such local
operator can (in principle) be written entirely in terms of currents (and products of currents) only. In other words, Diagram 2a and equation (\ref{D2a1}) generate not only corrections to the quadratic
kernel $f^{(2)}(q)$ but also to the cubic and higher kernels as well. In this paper we are only interested in the leading ${\cal O}(\bar q^2)$ correction
to $f^{(2)}(q)$ which corresponds to the $\bar q^2 (\bar\vf T^a\vf)_{q}$-term in the operator product expansion of (\ref{D2a3}). Therefore we will concentrate on extracting this term from (\ref{D2a3}).  

The integral $\Pi(r,q)$ admits a power series expansion in both arguments, so we may write
\beq
\Pi(r,q) = A \bar q^2 + B \bar q \bar r + C \bar r^2 + \ldots
\label{B6}
\eeq
where $A, B$ and $C$ are some constant coefficients\footnote{Numerical values of these constants can certainly be found from (\ref{D2a2}). However as will be seen shortly,
we do not really need to do that.} and the ellipsis stands for higher order terms (like $\bar q^2 \bar r q, \ \bar r^3 r$ {\it etc.}) in the expansion. Note that these 
higher order terms have {\it at least one} power of holomorphic momentum $q$ or $r$. Therefore, simple power counting suggests that such terms will
not contribute to the $\bar q^2 J^a_q$-term in the OPE of (\ref{D2a3}).

We will consider the contribution (to (\ref{D2a3})) of each of the three terms on the right hand side of
(\ref{B6}) separately.
The A-term is trivially evaluated as
\beq
\int \frac{d^2 r}{(2\pi)^2}\ [\bar\vf(r) T^a\vf(-r+q)+ \bar\vf(-r+q) T^a\vf(r)]\: A \bar q^2 =  2 ~A~ \bar q^2 J^a_q
\label{B7}
\eeq 
Similarly, the B-term can be evaluated as
\beqar
\int \frac{d^2 r}{(2\pi)^2}\ [\bar\vf(r) T^a\vf(-r+q)+ \bar\vf(-r+q) T^a\vf(r)]\: B \bar q \bar r &=& B \bar q \left[(\bar\partial\bar\vf T^a\vf)_{q}
+ (\bar\vf T^a\bar\partial\vf)_{q}\right] \nonumber\\
&=&B ~\bar q^2 J^a_q\label{B8}
\eeqar
The term involving $C {\bar r}^2$ requires more care.
Notice that by a change of variables ${\bar r} = -{\bar k} +{\half} {\bar q}$
in the first operator product and ${\bar r} = {\bar k} +{\half} {\bar q}$ in the second one we may write
\beqar
&&\int \frac{d^2 r}{(2\pi)^2}\ [\bar\vf(r) T^a\vf(-r+q)+ \bar\vf(-r+q) T^a\vf(r)]~ \bar r^2
\nonumber\\
&&= \int \frac{d^2 k}{(2\pi)^2}\ \bar\vf( -k +q/2) T^a\vf(k+q/2)\left[
({\bar k} + {\bar q}/2 )^2 + ( {\bar k} - {\bar q}/2)^2\right]
\label{B9}
\eeqar
For the composite operator $\bar\vf( -k +q/2) T^a\vf(k+q/2)$, $q$ is the total momentum and $2k$ is the relative momentum of the two fields
${\bar \vf}$ and $\vf$. Notice that powers of $k$ generate terms which are asymmetric between
$\vf$ and ${\bar \vf}$ in coordinate space, such as
${\bar \vf} \overleftrightarrow{\del}  T^a\vf$; such operator structures are not of
interest to us at present.
The term of interest corresponds to $k=0$, of the form
 ${\half} {\bar q}^2 \bar\vf(q/2) T^a\vf(q/2)$.
In other words, we may approximate
\beqar
&&\int \frac{d^2 r}{(2\pi)^2}\ [\bar\vf(r) T^a\vf(-r+q)+ \bar\vf(-r+q) T^a\vf(r)]\: C \bar r^2 \nonumber\\
&&\hspace{1in}\approx C \left[(\bar\partial^2\bar\vf T^a\vf)_{q}
+ (\bar\vf T^a\bar\partial^2\vf)_{q}\right]
= \frac{C}{2} \bar q^2 J^a_q
\label{B10}
\eeqar
Since 
\beq
2~A {\bar q}^2 +  B~{\bar q}^2 + {\half} C {\bar q}^2
= 2 (A {\bar q}^2 + B {\bar q} {\bar r} + C {\bar r}^2 )\biggr]_{{\bar r} = {\half} {\bar q}}
\label{B11}
\eeq
we may simplify the contribution (\ref{D2a1}) of Diagram 2a as
\beq
{\rm Diagram\ 2a} = \frac{2\pi}{m c_A} \int d\mu (q) (\bar\vf T^a\vf)_{-q} (\bar\vf T^a\vf)_{q}\ 2~\Pi(q/2,q) + \cdots
\label{B12}
\eeq
which in turn may be directly compared to (\ref{BF2}) to give
\beq
\Delta f^{(2)}(q) = 2\Pi(q/2,q) = \int \frac{d^2 k}{32\pi}\ \frac{g^{(3)}(k,q,-k-q)}{\bar q/2 + \bar k}\ \frac{m^2}{E_k\, E_{k+q}}
\label{B13}
\eeq
Computation of the leading ${\cal O}({\bar q}^2)$ term in the above integral is straightforward and can be done {\it analytically} to obtain
\beq
\Delta f^{(2)}(q) = \frac{\bar q^2}{2 m}\left( -\frac{1}{8} + 9\: {\rm log}\, 2 - 6\: {\rm log}\, 3\right) + \ldots = \frac{\bar q^2}{2 m}\,(-0.478349) +\ldots
\label{B14}
\eeq

\subsection*{Diagram 3}

Computation of Diagram 3 is quite similar to Diagram 2a which we have discussed at some length in the previous section. Straightforward contractions
dictated by Feynman graph in Figure 7 lead to the following expression
\beq\label{D31}
{\rm Diagram\ 3} = \frac{2\pi}{m c_A} \int \frac{d^2 q\, d^2 r}{(2\pi)^4}\ (\bar\vf T^a\vf)_{-q}\:[\bar\vf(r) T^a\vf(-r+q)+ \bar\vf(-r+q) T^a\vf(r)]\: \Pi(r,q)
\eeq
where now by $\Pi (r,q)$ we mean the following
\beq
\Pi (r,q) = \int \frac{d^2 k\, d^2 l}{2 (16\pi)^2}\ \frac{g^{(4)}(q, k ; l, -q-k-l)}{(\bar r + \bar k)(\bar r +\bar k +\bar l)}
\ \frac{m^3}{E_k\, E_l\, E_{k+l+q}}
\label{B15}
\eeq
We see that equation (\ref{D31}) is essentially identical to (\ref{D2a1}) and therefore all arguments from previous section apply to Diagram $3$ as well. In particular,
the correction to quadratic kernel $f^{(2)}(q)$ is 
\beq\label{D32}
\Delta f^{(2)}(q) =\ 2\,\Pi(q/2,q) = \int \frac{d^2 k\, d^2 l}{(16\pi)^2}\ \frac{g^{(4)}(q, k ; l, -q-k-l)}{(\bar q/2 + \bar k)(\bar q/2 +\bar k +\bar l)}
\ \frac{m^3}{E_k E_l E_{k+l+q}}
\eeq
The only difference from the previously considered example is that now we have to evaluate a two-loop integral which is difficult to do analytically.
However, it is quite easy to evaluate (\ref{D32}) {\it numerically} and we find
\beq
\Delta f^{(2)}(q) = \frac{\bar q^2}{2 m}(-0.23569 \pm 0.00001) + \ldots
\label{B16}
\eeq

\subsection*{Diagram 4a}
When evaluating Diagram 4a we will find that it is proportional to the following color factor
\beqar
f^{a_1 a_2 a_3} f^{b_1 b_2 b_3}
 \Tr(T^{a_2}T^{b_2}T^{a_3}T^{b_3})
&=& {\half} f^{a_1 a_2 a_3} f^{b_1 b_2 b_3} \Tr\bigl( \{ T^{a_2},T^{b_2}\} [T^{a_3},T^{b_3}] \nonumber\\
&&\hspace{.8in}+ [T^{a_2},T^{b_2}] \{T^{a_3},T^{b_3}\}\bigr) \nonumber\\ 
&\equiv&  0
\label{B17}
\eeqar
since the symmetric tensor $d^{abc} = \Tr (\{T^a, T^b\} T^c)=0$
for the adjoint representation.
Therefore we conclude that, by virtue of the color contractions, 
\beq
{\rm Diagram\ 4a} \equiv 0
\label{B18}
\eeq

\subsection*{Diagram 4b}
Once we do all bosonic line contractions as prescribed by Feynman graph in Figure 8, we obtain the following expression
\beq\label{D4b1}
\frac{2\pi}{m c_A} \int \frac{d^2 q}{(2\pi)^2}\ (\bar\vf T^a\vf)_{-q} (\bar\vf T^a\vf)_{q}\left[-\frac{\pi^3}{64 m} \int d\mu (k,l,r)
\ \frac{g^{(3)}(q,k,-q-k)\: g^{(3)}(q,l,-q-l)}{\bar r (\bar r -\bar q )(\bar r + \bar k)(\bar r + \bar l)}\right]
\eeq
Notice that the expression in square brackets is a three-loop integral that we have to evaluate. One of the loop integrals can be 
done {\it analytically}, namely,
\beq
\int \frac{d^2 r}{(2\pi)^2}\ \frac{1}{\bar r (\bar r -\bar q )(\bar r + \bar k)(\bar r + \bar l)} =
\frac{1}{\pi}\left[\frac{k}{\bar k (\bar k + \bar q)(\bar l -\bar k)} + \frac{l}{\bar l (\bar l + \bar q)(\bar k -\bar l)} - 
\frac{q}{\bar q (\bar k + \bar q)(\bar l +\bar q)}\right]
\label{B19}
\eeq
Using this result as well as inserting $4$ appropriate $m/E$ factors into (\ref{D4b1}) we end up with 
\begin{equation}
\begin{array}{rcl}
{\rm Diagram\ 4a} &=&\displaystyle \vspace{0.2in} \frac{2\pi}{m c_A} \int \frac{d^2 q}{(2\pi)^2}\ (\bar\vf T^a\vf)_{-q}\: (\bar\vf T^a\vf)_{q}\left\{ -\int \frac{d^2 k d^2 l}{(32\pi)^2}
\ \frac{m^3}{E_k E_l E_{k+q} E_{l+q}} \right .\\ \vspace{0.2in}
&&\displaystyle \hspace{0.6in} g^{(3)}(q,k,-q-k)\: g^{(3)}(q,l,-q-l) \left[ \frac{k}{\bar k (\bar k + \bar q)(\bar l -\bar k)}+ \right .\\
&&\displaystyle \hspace{0.6in} \left .\left . + \frac{l}{\bar l (\bar l + \bar q)(\bar k -\bar l)} - \frac{q}{\bar q (\bar k + \bar q)(\bar l +\bar q)}\right]\right\}
\end{array}
\label{B20}
\end{equation}
And this translates straightforwardly into the following correction to quadratic kernel $f^{(2)}$,
\begin{equation}
\begin{array}{rcl}
\Delta f^{(2)}(q)& =&\vspace{.2in}\displaystyle -\int \frac{d^2 k d^2 l}{(32\pi)^2}\ g^{(3)}(q,k,-q-k)\: g^{(3)}(q,l,-q-l)\ \frac{m^3}{E_k E_l E_{k+q} E_{l+q}}\\
&&\displaystyle
\hspace{0.5in} \times\ \left[\frac{k}{\bar k (\bar k + \bar q)(\bar l -\bar k)} + \frac{l}{\bar l (\bar l + \bar q)(\bar k -\bar l)} - 
\frac{q}{\bar q (\bar k + \bar q)(\bar l +\bar q)}\right] 
\end{array}
\label{B21}
\end{equation}
This (two-loop) integral can be easily evaluated {\it numerically} and  we find
\beq
\Delta f^{(2)}(q) = \frac{\bar q^2}{2 m}\: (0.020828 \pm 0.000002)\ +\ \ldots
\label{B22}
\eeq

\subsection*{Diagram 4c}

For Diagram\footnote{We consider Diagrams $4c$ and $4d$ as being different because we chose not to symmetrize completely the quartic vertex $F^{(4)}$.
Had we chosen to work with a completely symmeterized form of $F^{(4)}$, both diagrams would have been treated identically.} $4c$ we easily obtain
\beqar\label{D4c1}\nonumber
{\rm Diagram \ 4c}\!\! &=&\!\! \frac{2\pi}{m c_A} \int d\mu (q,r,s)\ [\bar\vf(r) T^a\vf(-r-q)+ \bar\vf(-r-q) T^a\vf(r)]\: \Pi(r,s,q)\\
&& \hspace{1.5in} \times\ [\bar\vf(s) T^a\vf(-s+q)+ \bar\vf(-s+q) T^a\vf(s)]
\eeqar
where $\Pi(r,s,q)$ stands for
\beq
\Pi(r,s,q) = \int \frac{d^2 k\, d^2 l}{(64\pi)^2}\ \frac{g^{(4)}(k, q-k; l,-q-l)}{(\bar k + \bar r)(\bar l + \bar s)}
\ \frac{m^4}{E_k\, E_l\, E_{k-q}\, E_{l+q}}
\label{B23}
\eeq
Straightforward generalization of the analysis presented for Diagram $2a$ suggests that we should evaluate $\Pi(r,s,q)$ at $r = -q/2$ and $s = q/2$. In other words we
may write the relevant term of (\ref{D4c1}) as 
\beq
{\rm Diagram\ 4c} = \frac{2\pi}{m c_A} \int \frac{d^2 q}{(2\pi)^2} (\bar\vf T^a\vf)_{-q} (\bar\vf T^a\vf)_{q}\ ~4~\Pi(-q/2, q/2, q) + \cdots
\label{B24}
\eeq
from which we immediately conclude that
\beq\label{D4c4}
\Delta f^{(2)}(q) =\ 4\,\Pi(-q/2, q/2, q) = \int \frac{d^2 k\, d^2 l}{(32\pi)^2}\ \frac{g^{(4)}(k, q-k; l,-q-l)}{(\bar k - \bar q/2)(\bar l + \bar q/2)}
\ \frac{m^4}{E_k\, E_l\, E_{k-q}\, E_{l+q}}
\eeq
Th integral in (\ref{D4c4}) can be evaluated {\it numerically} and we find
\beq
\Delta f^{(2)}(q) = \frac{\bar q^2}{2 m}\: (-0.06893 \pm 0.00002) +\ \ldots
\label{B25}
\eeq

\subsection*{Diagram 4d}

The mathematical expression for diagram 4d splits into two parts involving symmetrized and antisymmetrized products of the color matrices. Thus
\beq
{\rm Diagram\ 4d} = [{\rm Diagram\ 4d}]_S + [{\rm Diagram\ 4d}]_A
\eeq
where
\beqar
[{\rm Diagram\ 4d}]_S\!\! &=&\!\! -\frac{e^2}{8} \left(\frac{2\pi}{m c_A}\right)^2 \!\!\! \int
d\mu (k_1, k_2, q_1, q_2, r, s) \ f^{(4)}_{a_1 a_2; b_1 b_2}(k_1, k_2; q_1, q_2)\nonumber\\
&& \quad \times\ [\bar\vf(r)\{ T^{a_1},T^{b_1}\}\left(\frac{1}{\bar r +\bar k_1} +\frac{1}{\bar r + \bar q_1}\right)\vf (-r-k_1-q_1)]\nonumber\\
&& \quad \times\ [\bar\vf(s)\{ T^{a_2},T^{b_2}\}\left(\frac{1}{\bar s +\bar k_2} +\frac{1}{\bar s + \bar q_2}\right)\vf (-s-k_2-q_2)]
\label{Diag4dS}
\eeqar
The color structure here is such that we will not generate a term of the current-current form.
If we expand the symmetrized product of the adjoint matrices, say,  $\{ T^{a_1}, T^{b_1}\}$
in terms of a basis of hermitian matrices, the term proportional to
$T^c$ will have the coefficient $ d^{a_1 b_1 c} = \Tr  \{ T^{a_1}, T^{b_1}\} T^c$, which is zero for the adjoint representation. Thus the contribution involving the current-current form comes from
the second part of the diagram with the antisymmetrized products. It is given by
\beqar
[{\rm Diagram\ 4d}]_A\!\! &=&\!\! -\frac{e^2}{8} \left(\frac{2\pi}{m c_A}\right)^2 \!\!\! \int d\mu (k_1, k_2, q_1, q_2, r, s) \ f^{(4)}_{a_1 a_2; b_1 b_2}(k_1, k_2; q_1, q_2)\nonumber\\
&& \quad \times\ [\bar\vf(r) [T^{a_1},T^{b_1}] \left(\frac{1}{\bar r +\bar k_1} -\frac{1}{\bar r + \bar q_1}\right)\vf (-r-k_1-q_1)]\nonumber\\
&& \quad \times\ [\bar\vf(s) [T^{a_2},T^{b_2}] \left(\frac{1}{\bar s +\bar k_2} -\frac{1}{\bar s + \bar q_2}\right)\vf (-s-k_2-q_2)]
\label{Diag4d1}
\eeqar
This expression can be further simplified as
\beqar
[{\rm Diagram \ 4d}]_A\!\! &=&\!\! \frac{2\pi}{m c_A} \int d\mu (q,r,s)\ [\bar\vf(r) T^a\vf(-r-q)+ \bar\vf(-r-q) T^a\vf(r)]\: \Pi(r,s,q)\nonumber\\
&& \hspace{1in} \times\ [\bar\vf(s) T^a\vf(-s+q)+ \bar\vf(-s+q) T^a\vf(s)]
\label{Diag4dA2}
\eeqar
where, we have once again,
\beq
\Pi(r,s,q) = \int \frac{d^2 k\, d^2 l}{(64\pi)^2}\ \frac{g^{(4)}(k,l; q-k; ,-q-l)}{(\bar k + \bar r)(\bar l + \bar s)}
\ \frac{m^4}{E_k\, E_l\, E_{k-q}\, E_{l+q}}
\label{Diag4dA3}
\eeq
The rest of the analysis is similar to the case of diagram 4c, giving the contribution
\beq
\Delta f^{(2)}(q) =\ 4\,\Pi(-q/2, q/2, q) = \int \frac{d^2 k\, d^2 l}{(32\pi)^2}\ \frac{g^{(4)}(k, l; q-k, -q-l)}{(\bar k - \bar q/2)(\bar l + \bar q/2)}
\ \frac{m^4}{E_k\, E_l\, E_{k-q}\, E_{l+q}}
\label{Diag4dA4}
\eeq
{\it Numerical} evaluation of the integral gives
\beq
\Delta f^{(2)}(q) = \frac{\bar q^2}{2 m}\: (-0.01216 \pm 0.00007) +\ \ldots
\label{Diag4dA5}
\eeq

\begin{figure}[!t]
\begin{center}
\includegraphics[height = .2\textwidth, width=.4\textwidth]{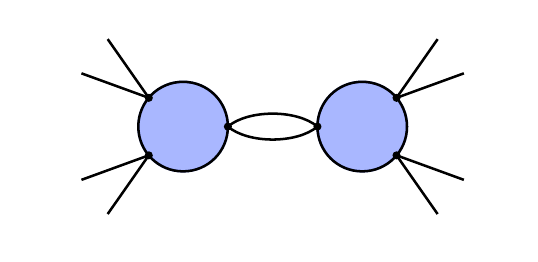}\\
\end{center}
\caption{``Effective" quartic vertex.}
\label{DiagEffective}
\end{figure}
\subsection*{Diagram $4e$}

The simplest way to compute Diagram $4e$ is to introduce an "effective" quartic vertex 
\beq
g^{(4)}_{eff}\,(k_1, k_2; q_1, q_2) =\ g^{(3)}(k_1, k_2, -k_1 - k_2)\ \frac{k_1 + k_2}{\bar k_1 + \bar k_2}\,\frac{1}{E_{k_1+k_2}}\ g^{(3)}(q_1, q_2, -q_1 - q_2)
\eeq 
as shown on Figure \ref{DiagEffective}, and to notice that Diagram $4e$ is quite similar to Diagram $3$. Therefore, we
may simply re-use previously derived equations (\ref{D31}-\ref{D32}) by replacing $g^{(4)}$ with $g^{(4)}_{eff}$. In this way
we immediately find
\beq
{\rm Diagram\ 4e} = \frac{2\pi}{m c_A} \int d\mu (q,r)\ (\bar\vf T^a\vf)_{-q}\:[\bar\vf(r) T^a\vf(-r+q)+ \bar\vf(-r+q) T^a\vf(r)] ~\Pi(r,q)
\label{D4e1}
\eeq
with 
\beq
\Pi (r,q) = \int \frac{d^2 k\, d^2 l}{2 (16\pi)^2}\ \frac{g^{(4)}_{eff}(q, k ; l, -q-k-l)}{(\bar r + \bar k)(\bar r +\bar k +\bar l)}
\ \frac{m^3}{E_k\, E_l\, E_{k+l+q}}
\label{Diag4e2}
\eeq
As usual, correction to $f^{(2)}(q)$ is 
\beq
\Delta f^{(2)}(q) =\ 2\,\Pi(q/2,q) = \int \frac{d^2 k\, d^2 l}{(16\pi)^2}\ \frac{g^{(4)}_{eff}(q, k ; l, -q-k-l)}{(\bar q/2 + \bar k)(\bar q/2 +\bar k +\bar l)}
\ \frac{m^3}{E_k\, E_l\, E_{k+l+q}}
\eeq
{\it Numerical} evaluation of this integral presents no difficulties and we find
\beq
\Delta f^{(2)}(q) = \frac{\bar q^2}{2 m}(-0.06824 \pm 0.00002) + \ldots
\eeq
\subsection*{Diagram $4f$}
Diagram 4f has the structure of the product of an effective vertex corresponding to diagram 2a 
and the $F^{(2)}$ vertex with two Wick contractions connecting them.
It can also be directly written down from the interaction terms.
We get
\beq
{\rm Diagram\ 4f} = \frac{2\pi}{m c_A} \int d\mu (q,r)\ (\bar\vf T^a\vf)_{-q}\:[\bar\vf(r) T^a\vf(-r+q)+ \bar\vf(-r+q) T^a\vf(r)] ~\Pi(r,q)
\label{D4f1}
\eeq
where
\beqar
\Pi (r,q) &=& \int {d^2k d^2l \over 256 \pi^2}
~{{\bar l}^2~g ^{(3)}(k, q, -k-q)\over ({\bar r} + {\bar l}) ( {\bar r} + {\bar l} - {\bar q})}
\left( {1\over {\bar r} + {\bar l} + {\bar k} } + {1\over {\bar q} +{\bar k} - {\bar r} - {\bar l}}
\right) \nonumber\\
&&\hskip .2in \times {m^2 \over (m+E_l) E_l E_k E_{k+q} }
\label{D4f2}
\eeqar
We can now follow a similar line of reasoning as we did for diagrams 2a, 3, 4c-4e and conclude
that
\beqar
\Delta f^{(2)} (q) &=& 2 ~\Pi ( q/2, q)\nonumber\\
&=&  \int {d^2k d^2l \over 128 \pi^2}~g ^{(3)}(k, q, -k-q)
~{ {\bar l}^2\over  ({\bar l}^2 - ({\bar q}/2)^2)}
\left( {1\over ({\bar q}/2) + {\bar l} + {\bar k} } + {1\over ({\bar q}/2) +{\bar k}  - {\bar l}}
\right) \nonumber\\
&&\hskip 1in \times {m^2 \over (m+E_l) E_l E_k E_{k+q} }
\label{D4f3}
\eeqar
The {\it numerical} evaluation of this integral gives
\beq
\Delta f^{(2)} (q) = {{\bar q}^2 \over 2m} ( - 0.1666 \pm 0.0002 )
\label{D4f4}
\eeq

This diagram, as explained in text, has some subtleties. We have argued that, to avoid double counting of a part of this diagram, we have to restrict the integration over
the second loop momentum $k$ to values above the low momentum cut off.
Taking the value of the cut off as $m$ and  $2m$ in turn,
we find
\beq
\Delta f^{(2)} (q) =  {{\bar q}^2 \over 2m} \left\{ \begin{matrix}
 ( - 0.1037 \pm 0.0003 )& \hskip .2in {\rm for~ cut off}~= m\\
( - 0.051104 \pm 0.00019 )& \hskip .2in {\rm for~ cut off}~= 2m\\
  \end{matrix} \right.
\label{D4f5}
\eeq


\end{document}